\documentclass[aps,pra,letterpaper,10pt,twocolumn,showpacs,superscriptaddress,notitlepage]{revtex4-2}

\usepackage[utf8]{inputenc}
\usepackage{amsfonts,amssymb,amsmath,amsthm}
\usepackage{url}
\usepackage{dsfont}
\usepackage{bbm}
\usepackage[dvipsnames]{xcolor}
\usepackage[table]{xcolor}
\usepackage{nicefrac}
\usepackage{setspace}
\usepackage{tabularx,booktabs}
\usepackage{makecell}
\usepackage{comment}
\usepackage[normalem]{ulem}
\usepackage{siunitx}
\usepackage{physics}
\usepackage{comment}
\usepackage{graphicx}
\usepackage{multirow}

\usepackage[colorlinks=true, citecolor=blue,urlcolor=blue,linkcolor=blue,filecolor=black]{hyperref}
\renewcommand{\selectlanguage}[1]{}

\usepackage{siunitx}
\usepackage{ragged2e}

\makeatletter
\renewcommand{\fnum@figure}{\justifying\figurename~\thefigure}
\renewcommand{\fnum@table}{\justifying\tablename~\thetable}
\makeatother

\DeclareSIUnit{\pulse}{pulse}
\DeclareSIUnit{\sample}{Sa}
\DeclareSIUnit{\snu}{SNU}
\DeclareSIUnit{\symbol}{symbol}
\DeclareSIUnit{\baud}{Baud}
\DeclareSIUnit{\sample}{S}

\usepackage{glossaries-extra}
\setabbreviationstyle[acronym]{long-short}
\glsdisablehyper
\newacronym{qrng}{QRNG}{quantum random number generator}
\newacronym{qkd}{QKD}{quantum key distribution}
\newacronym{dv}{DV}{discrete variables}
\newacronym{cv}{CV}{continuous variables}
\newacronym{pic}{PIC}{photonic integrated circuit}
\newacronym{plc}{PLC}{planar waveguide circuit}
\newacronym{soi}{SOI}{silicon-on-insulator}
\newacronym{cmos}{CMOS}{complementary metal-oxide-semiconductor}
\newacronym{flm}{FLM}{femtosecond laser micromachining}
\newacronym{cots}{COTS}{commercial off-the-shelf}
\newacronym{mzi}{MZI}{Mach-Zehnder interferometer}
\newacronym{dc}{DC}{directional coupler}
\newacronym{tops}{TOPS}{thermo-optic phase shifter}
\newacronym{lo}{LO}{local oscillator}
\newacronym{bpd}{BPD}{balanced photodetector}
\newacronym{ecl}{ECL}{external-cavity laser}
\newacronym{pbs}{PBS}{polarizing beam splitter}
\newacronym{pc}{PC}{polarization controller}
\newacronym{fpga}{FPGA}{field-programmable gate array}
\newacronym{qpsk}{QPSK}{quadrature phase-shift keying}
\newacronym{psk}{PSK}{phase-shift keying}
\newacronym{qam}{QAM}{quadrature amplitude modulation}
\newacronym{awg}{AWG}{arbitrary waveform generator}
\newacronym{hpf}{HPF}{high-pass filter}
\newacronym{att}{ATT}{attenuator}
\newacronym{voa}{VOA}{variable optical attenuator}
\newacronym{bs}{BS}{beam splitter}
\newacronym{pm}{PM}{power meter}
\newacronym{cmrr}{CMRR}{common-mode rejection ratio}
\newacronym{fwhm}{FWHM}{full width at half maximum}
\newacronym{povm}{POVM}{positive operator-valued measure}
\newacronym{skr}{SKR}{secret key rate}
\newacronym{rrc}{RRC}{root-raised-cosine}
\newacronym{pcs}{PCS}{probabilistic constellation shaping}
\newacronym{tia}{TIA}{transimpedance amplifier}
\newacronym{vbs}{vBS}{variable beam splitter}
\newacronym{iso}{ISO}{isolator}
\newacronym{dsp}{DSP}{digital signal processing}
\newacronym{ft}{FT}{fully trusted}
\newacronym{di}{DI}{device-independent}
\newacronym{sdi}{SDI}{semi-device-independent}
\newacronym{eat}{EAT}{entropy accumulation theorem}
\newacronym{ook}{OOK}{on-off keying}
\newacronym{bpsk}{BPSK}{binary phase-shift keying}
\newacronym{sdp}{SDP}{semidefinite programming}

\usepackage{todonotes}
\definecolor{marcocol}{HTML}{5366ff}
\definecolor{armincolbkg}{HTML}{5366ff}

\begin{document}

\title{Efficient Energy-Constrained Semi-Device-Independent QRNG with an Integrated Heterodyne Receiver}

\author{Mattia Sabatini}
\thanks{These two authors contributed equally}
\affiliation{Dipartimento di Ingegneria dell’Informazione, Università degli Studi di Padova, Via Gradenigo 6B, 35131 Padova, Italy}

\author{Carles Roch i Carceller}
\thanks{These two authors contributed equally}
\affiliation{Physics Department and NanoLund, Lund University, Box 118, 22100 Lund, Sweden}

\author{Tommaso Bertapelle}
\affiliation{Dipartimento di Ingegneria dell’Informazione, Università degli Studi di Padova, Via Gradenigo 6B, 35131 Padova, Italy}

\author{Andrea Peri}
\affiliation{Dipartimento di Ingegneria dell’Informazione, Università degli Studi di Padova, Via Gradenigo 6B, 35131 Padova, Italy}

\author{Giulio Gualandi}
\affiliation{Dipartimento di Fisica, Politecnico di Milano, Piazza Leonardo da Vinci 32, 20133 Milano, Italy}
\affiliation{Istituto di Fotonica e Nanotecnologie, Consiglio Nazionale delle Ricerche (CNR), Piazza Leonardo da Vinci 32, 20133 Milano, Italy}

\author{Matías Rubén Bolaños}
\affiliation{Dipartimento di Ingegneria dell’Informazione, Università degli Studi di Padova, Via Gradenigo 6B, 35131 Padova, Italy}

\author{Yoann Piétri}
\affiliation{Dipartimento di Ingegneria dell’Informazione, Università degli Studi di Padova, Via Gradenigo 6B, 35131 Padova, Italy}

\author{Giacomo Corrielli}
\affiliation{Istituto di Fotonica e Nanotecnologie, Consiglio Nazionale delle Ricerche (CNR), Piazza Leonardo da Vinci 32, 20133 Milano, Italy}

\author{Paolo Villoresi}
\affiliation{Dipartimento di Ingegneria dell’Informazione, Università degli Studi di Padova, Via Gradenigo 6B, 35131 Padova, Italy}
\affiliation{Padua Quantum Technologies Research Center, Università degli Studi di Padova, via Gradenigo 6B, 35131 Padova, Italy}

\author{Giuseppe Vallone}
\affiliation{Dipartimento di Ingegneria dell’Informazione, Università degli Studi di Padova, Via Gradenigo 6B, 35131 Padova, Italy}
\affiliation{Padua Quantum Technologies Research Center, Università degli Studi di Padova, via Gradenigo 6B, 35131 Padova, Italy}

\author{Roberto Osellame}
\affiliation{Istituto di Fotonica e Nanotecnologie, Consiglio Nazionale delle Ricerche (CNR), Piazza Leonardo da Vinci 32, 20133 Milano, Italy}

\author{Armin Tavakoli}
\affiliation{Physics Department and NanoLund, Lund University, Box 118, 22100 Lund, Sweden}

\author{Marco Avesani}
\email[Corresponding email: ]{marco.avesani@unipd.it}
\affiliation{Dipartimento di Ingegneria dell’Informazione, Università degli Studi di Padova, Via Gradenigo 6B, 35131 Padova, Italy}
\affiliation{Padua Quantum Technologies Research Center, Università degli Studi di Padova, via Gradenigo 6B, 35131 Padova, Italy}

\begin{abstract}
Semi-device-independent QRNG frameworks represent a particularly attractive approach, combining strong security guarantees with high randomness generation rates while relying only on reduced and practical physical assumptions.
A recently proposed approach based on photon-number constraints is particularly suited to photonic implementations, where these assumptions can be easily assessed experimentally. Here, we experimentally demonstrate a quantum random number generator within this framework, enabling the direct computation of lower bounds on the certifiable Shannon entropy via semidefinite relaxation techniques. When combined with entropy accumulation methods, this approach enables finite-size randomness certification without assuming independent and identically distributed rounds. We realize the protocol using a four-state coherent-state constellation symmetrically distributed in phase space and measured by heterodyne detection, certifying 0.223 bit per measurement, which is the highest value reported to date for a continuous-variable semi-device-independent QRNG. The implementation combines a low-loss integrated photonic heterodyne receiver with a simple transmitter assembled from commercial components, providing a practical and high-speed architecture for semi-device-independent randomness generation.
\end{abstract}

\maketitle

\section{Introduction}
\Glspl{qrng} are among the most mature technologies in quantum information science, developed to deliver genuine uncertainty.
Among its applications, true randomness plays a central role in modern cryptography since it is necessary to ensure security.
With \glspl{qrng}, quantum physics enables randomness certification directly from fundamental physical laws~\cite{Acin2016}.
Depending on the working hypothesis used to certify randomness, \glspl{qrng} are classified in \gls{ft}, \gls{di}, and \gls{sdi}: the first requires full device characterization~\cite{Pan2015, Bruynsteen2023}, the second operates without trusting its implementation~\cite{Pironio2010,bierhorst_experimentally_2018,liu_device-independent_2018,li_experimental_2021,liu_device-independent_2021}
, and the third
relies on relaxing the DI approach through a mild assumption about the system; see for instance the review~\cite{Brask2026}.
Although fully \gls{di} protocols offer the strongest security guarantees, they are experimentally challenging and their achievable randomness generation rates  still limited for several practical use cases.. Consequently, \gls{sdi} protocols, which require only minor knowledge of the devices, have been extensively studied as a pragmatic alternative, combining high generation rates with reduced implementation complexity~\cite{Cao2015, Avesani2018, Marangon2019, Brask2017, Avesani2021}.
Early SDI-approaches to QRNG relied on bounding the Hilbert space dimension of the system~\cite{Li2011, Li2012}. While conceptually appealing, this assumption has notable drawbacks: i) it is commonly associated with single photons which have low generation rates, ii) the dimension is not an observable quantity and it is therefore difficult to verify in experiment, and iii) finite-dimensional implementations, typically qubits, are usually approximations of what is actually higher-dimensional systems and small such components can undermine the security of QRNG~\cite{Pauwels2022}.
To overcome these obstacles, alternative \gls{sdi} frameworks have been developed in recent years, proposing to deduce the security of \gls{qrng} from more practically motivated assumptions. For instance, \glspl{qrng} have been designed based on assuming the overlap between pure state preparations~\cite{Brask2017, Tebyanian2021_2nd}.  However, overlap-based assumptions are best suited to pure-state preparations, while their interpretation becomes less direct for mixed states. A more practical alternative is to use an energy-based assumption, as done in Ref.~\cite{Avesani2021}, where the constraint is expressed as a bound on the average photon number, or equivalently on the vacuum component, of the prepared states. This constraint is physically motivated and experimentally accessible, since it can be estimated by monitoring the optical energy emitted by the source. In the i.i.d.~regime, this constraint can be applied to each preparation independently. However, cryptographic applications often require certification without this assumption. In the resulting non-i.i.d.~regime, constraints on averaged observable quantities provide a more suitable description.
A framework based on such average constraints was first introduced in~\cite{VanHimbeeck2017, vanhimbeeck2019} and then experimentally realized in~\cite{rusca_fast_2020} with homodyne detection. This method is limited to two coherent state preparations that form a \gls{bpsk} constellation. However, the recent theoretical work of Ref.~\cite{carceller2025} has developed methods to enable the analysis of randomness generated from more than two coherent states and measurements with multiple outcomes, thereby providing a path to enhanced \gls{qrng} protocols and implementations.

{In this work, we develop and implement a \gls{qrng} protocol based on the framework of Ref.~\cite{carceller2025}, allowing us to operate, for the first time, beyond \gls{bpsk} while incorporating finite-size effects in a non-i.i.d.~regime.
Indeed, with the semidefinite relaxation techniques used in Ref.~\cite{carceller2025}, the protocol lower-bounds the smooth conditional min-entropy $H_{\rm min}^\varepsilon(X|E)$ by directly bounding the conditional Shannon entropy. This leads to a stronger randomness certification per measurement than the conditional min-entropy bounds often used in the literature.}
Our implementation employs a \gls{qpsk} phase-space constellation of four symmetrically distributed coherent states measured with optical heterodyne detection.
When aiming towards real-world applications, not only performance but also considerations such as size, cost, system integration, scalability, reliability, and power consumption are required. For this reason, our implementation is realized using a stable, high-performance, and low-loss glass-based integrated photonic chip highly compatible with standard \SI{1550}{\nano\meter} fiber-based telecom components~\cite{Peri2025}.
This platform allows us to pair such a receiver with a simple \gls{qpsk} transmitter assembled from \gls{cots} components~\cite{Sabatini2025}.
Together with the theoretical framework of Ref.~\cite{carceller2025}, we were able to certify with our telecom-compatible integrated coherent receiver up to \SI{0.223}{bit} per measurement in a finite-size non~i.i.d. scenario via the \gls{eat}~\cite{metger2022}. At a repetition rate of \SI{250}{\mega baud}, this corresponds to a secure generation rate of \SI{55.6}{\mega bit/\second}. To the best of our knowledge, this is the highest certified entropy per measurement reported among coherent-detection implementations of this class of \gls{sdi}-\gls{qrng}.

The manuscript is structured as follows. In Section~\ref{sec:protocol} we introduce the \gls{sdi} framework and the numerical method used to perform the randomness certification; in Section~\ref{sec:exp_setup} we describe the experimental setup adopted. In Section~\ref{sec:results} we report the results obtained, and conclude in Section~\ref{sec:Conclusion}.

\section{Semi-DI QRNG protocol}\label{sec:protocol}
\begin{figure}[htbp!]
    \includegraphics[width=0.42\textwidth]{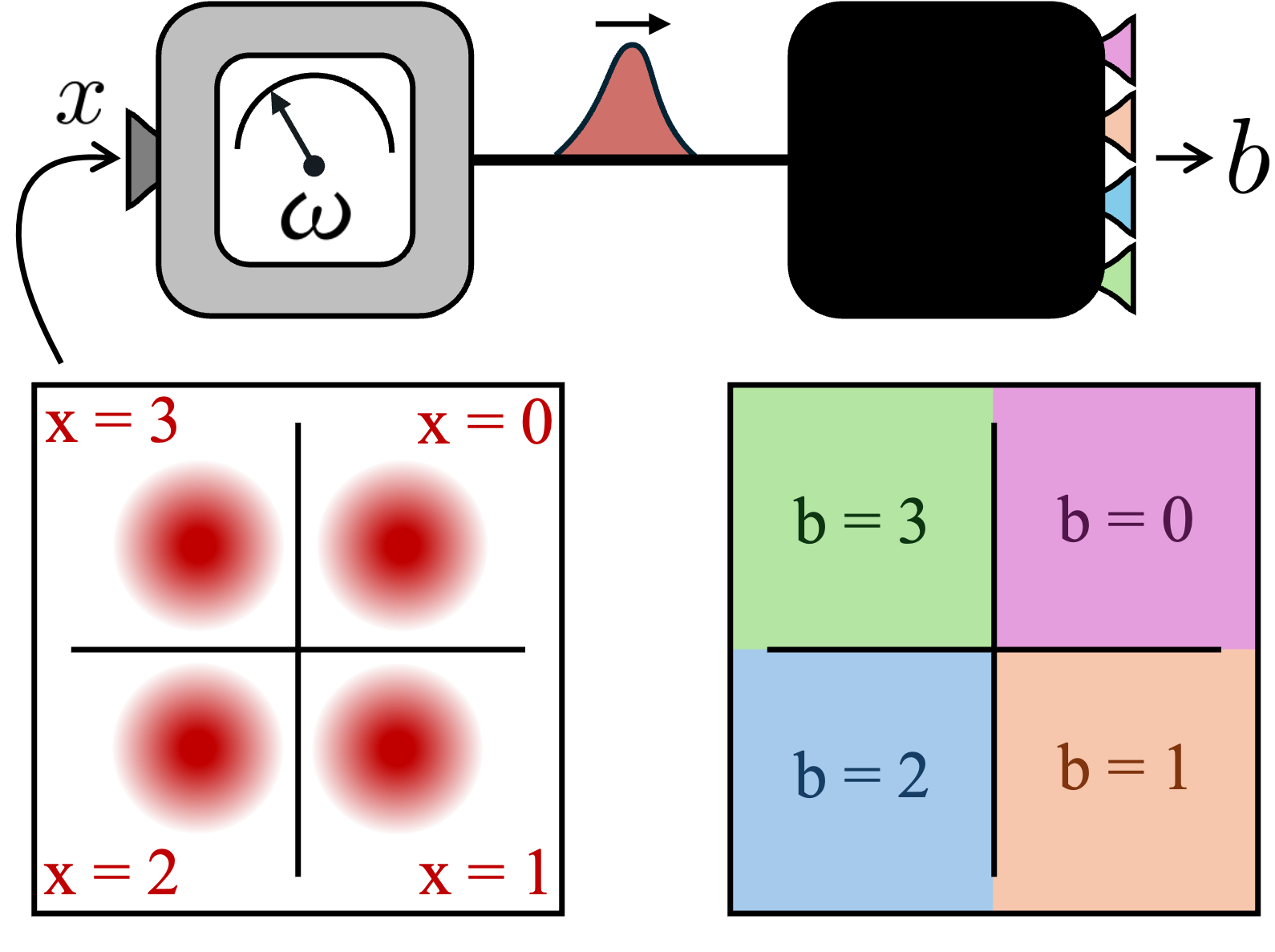}
    \caption{\textbf{Prepare-and-measure protocol.} A device prepares a coherent state according to the input $x\in\{0,1,2,3\}$. This state is sent to a second device that performs a heterodyne measurement with outcome $b \in \{0,1,2,3\}$ corresponding to the four quadrants in phase-space.}
    \label{fig:PAM}
\end{figure}

\subsection{Prepare-and-measure framework}
Our \gls{qrng} protocol operates in the prepare-and-measure scenario; see Fig.~\ref{fig:PAM}. A transmitter (Alice) receives a private quaternary input $x\in\{0,1,2,3\}$ with probability $p_x$ and prepares a corresponding quantum state $\rho_x$. This state is sent to the receiver (Bob), which performs a measurement described by the \gls{povm} $\{M_b\}$, producing a measurement outcome \(B\) with realizations \(b\in\{0,1,2,3\}\). We also allow Alice and Bob to use different preparation and measurement strategies in each round, labeled by a  shared hidden variable $\lambda$ drawn from a distribution $q(\lambda)$, that models shared classical correlation between the parties. The observable input-output probabilities averaged over the strategies $\lambda$ are $p(b|x)=\sum_\lambda q(\lambda)p_\lambda(b|x)$, where the correlations per-round $p_\lambda(b|x)=\tr\small(\rho_x^{(\lambda)} M_b^{(\lambda)}\small)$ are given by the Born rule. We consider that both the preparation and measurement devices are fully uncharacterized, with the exception of the following \gls{sdi} assumption: the average vacuum-component of the states is lower bounded. This is enforced by the condition
\begin{align}\label{eq:sdi_ass}
\sum_{x} p_x \bra{0} \rho_x \ket{0} \geq \omega \ ,
\end{align}
where $\rho_x = \sum_\lambda q(\lambda) \rho_x^{(\lambda)}$ is the average state and $\omega$ is a lower bound on the measurable probability that the preparation device emits zero photons.

Our goal is to certify the randomness of Bob's measurement outcome $B$.
The certification must be achieved with respect to an eavesdropper (Eve) whose objective is to guess $b$.
As in Ref.~\cite{carceller2025}, we consider an adversary with classical side information: Eve has access to the hidden variable \(\lambda\), which classically correlates Alice's and Bob's devices, but holds no quantum side information entangled with them. This assumption is well motivated for a \gls{qrng}, where the device can in practice be confined to a locally operated chip, making quantum side information less relevant than in scenarios such as \gls{qkd}, where an adversary may interact with signals propagating through an external quantum channel.

To limit Eve's guessing capabilities, the prepare-and-measure scenario is designed to satisfy a bound on a particular linear correlation. Concretely, we constrain the state discrimination success probability---i.e.~the probability that Bob's outcome matches Alice's input---to be lower-bounded by an observed value $W$ as
\begin{align}\label{eq:witness}
\sum_{x}p_x \sum_\lambda q(\lambda) p_\lambda (x|x) \geq W \ .
\end{align}
Any eavesdropping strategy determined by $\lambda$ must then, on average, satisfy the bound on the observable correlation above.

\subsection{Protocol implementation}
We implement the prepare-and-measure scenario in the lab using four coherent state preparations 
$\{\ket{\alpha_x}\}_x$, with complex amplitudes $\alpha_x = \abs{\alpha}e^{i\frac{(1-2x)\pi}{4}}$. These states are symmetrically distributed around the vacuum in the phase-space, as shown in Fig.~\ref{fig:PAM}, and contain the same vacuum component $\omega = e^{-\abs{\alpha}^2}$, where $\abs{\alpha}^2$ corresponds to the averaged photon-number experimentally measured.
Even without assuming a specific photon-number distribution, any quantum state with mean photon number $\abs{\alpha}^2<1$ must have a nonzero vacuum component. This gives the distribution-independent lower bound
\begin{align}
\omega \geq 1-\abs{\alpha}^2 .
\end{align} This bound therefore only requires knowledge of the mean photon number $\abs{\alpha}^2$. In practice, this quantity can be directly estimated with a power meter, so that the physical assumption remains experimentally accessible.

In each prepare-and-measure round, the prepared coherent state is sent to a device that performs a heterodyne measurement.
By associating each outcome \(\beta=\beta_R+i\beta_I\) with one of the four phase-space quadrants, we obtain the effective four-outcome \gls{povm}:
\begin{equation}  
M_b=\int_{Q_b}\frac{d^2\beta}{\pi}\,\ket{\beta}\!\bra{\beta},
\qquad b\in\{0,1,2,3\},
\end{equation}
with:
\begin{equation}
    \begin{split}
        Q_0 &= \{\beta\in\mathbb{C}:\operatorname{Re}\beta\ge 0,\ \operatorname{Im}\beta\ge 0\}\\
        Q_1 &= \{\beta\in\mathbb{C}:\operatorname{Re}\beta\ge 0,\ \operatorname{Im}\beta< 0\}\\
        Q_2 &= \{\beta\in\mathbb{C}:\operatorname{Re}\beta< 0,\ \operatorname{Im}\beta< 0\}\\
        Q_3 &= \{\beta\in\mathbb{C}:\operatorname{Re}\beta< 0,\ \operatorname{Im}\beta\ge 0\}\\
    \end{split}\, ,
\end{equation}
With that, we compute the observable input-output probabilities $p(b|x)=\frac{1}{\pi}\int_{Q_b}d^2\beta \ \abs{\braket{\alpha_x}{\beta}}^2$ and estimate the security parameter to be:
\begin{align}
    W = \frac{1}{4}\bigg{[} 1 + \text{erf}\left(\frac{\abs{\alpha}}{\sqrt{2}}\right) \bigg{]}^2 \ ,
\end{align}
where $\text{erf}(\cdot)$ is the error function. 

\subsection{Randomness certification}
In the asymptotic regime, where a prepare-and-measure experiment is repeated over many independent and identically distributed rounds, the randomness generated in the measurement outcomes is quantified by the conditional von Neumann entropy $H(\mathbf{B}|\mathbf{E})$ of the full sequence of $n$ measurement outcomes $\mathbf{B} = (B_1, \ldots, B_n)$, conditioned on all side information $\mathbf{E}$ available to an adversary. However, in realistic randomness certification protocols, where the devices may be imperfect or even corrupted by an adversary, such an i.i.d. assumption can be overly restrictive, as it prevents the adversary from exploiting correlations across different rounds.

For this reason, it is essential to consider the more general finite-size, non-i.i.d.~regime, which constitutes the setting of this work. In this case, the relevant figure of merit is the smooth conditional min-entropy $H_{\min}^\varepsilon(\mathbf{B}|\mathbf{E})$, which quantifies the amount of randomness that can be securely extracted from the measurement's outcomes ~\cite{Renner05}. Operationally, this quantity determines the number of nearly uniform and independent random bits that can be obtained after classical post-processing via randomness extractors.

The central task in randomness certification is therefore to obtain reliable lower bounds on $H_{\min}^\varepsilon(\mathbf{B}|\mathbf{E})$. In the asymptotic limit, this quantity converges to the single-round conditional von Neumann entropy. For our eavesdropper with side-information $\Lambda_E$, this  reduces to the conditional Shannon entropy \cite{carceller2025}, 
\begin{align}
H(B|\Lambda_E) = -\sum_x p_x \sum_\lambda q(\lambda) p_\lambda(b|x) \log_2(p_\lambda(b|x)) \ .
\end{align}
The nonlinear form of the Shannon entropy makes the certification of valid lower bounds a challenging task. While a simple bound can be obtained by considering the non-smooth min-entropy via \gls{sdp} relaxation methods~\cite{SDPreview}, this usually leads to a significant underestimation of the Shannon entropy. 
Following Ref.~\cite{carceller2025}, however, \gls{sdp} relaxation methods can instead be formulated to bound directly the Shannon entropy.

More specifically, this approach is based on \gls{sdp} relaxations of the set of quantum correlations arising under restrictions on the photon-number components. The \gls{sdi} assumption from \eqref{eq:sdi_ass} can then be directly implemented by limiting the zero-photons (vacuum) component averaged over Alice's inputs. We refer to Appendix \ref{app.sdp_shannon} for details on the \gls{sdp} relaxations used to bound the Shannon entropy. Once the primal \gls{sdp} is solved, its corresponding dual formulation can be derived explicitly. Formulating the \gls{sdp} in its dual form is both relevant and meaningful to certify the extractable randomness from an experimental implementation in a twofold manner. Firstly, experimental data provide empirical frequencies that, due to finite-size effects, only approximate the underlying probability distribution. Consequently, the estimated distribution may not admit a valid quantum realization, in which case the constraints of the primal \gls{sdp} cannot be satisfied exactly, rendering the problem infeasible. In contrast, since the optimal value of the dual always yields an upper-bound on the primal solution, the dual program remains feasible even when the observed correlations are not physically realizable. Secondly, because the \gls{sdp} relaxation incorporates the linear constraints associated with \eqref{eq:sdi_ass} and \eqref{eq:witness}, the dual objective function is affine and depends linearly on both the observed witness value $W$ and the bound on the vacuum component $\omega$. 

To extend the randomness certification to the full sequence in the non-i.i.d.~regime, we employ the Entropy Accumulation Theorem~(EAT)~\cite{metger2022}, which lower-bounds $H_{\min}^\varepsilon(\mathbf{B}|\mathbf{E})$ in terms of a \emph{min-tradeoff function} $f_{\min}$, i.e.~an affine function on the observable correlation that lower-bounds the single-round Shannon entropy, as
\begin{align}
    \frac{1}{n}H_{\min}^\varepsilon(\mathbf{B}|\mathbf{E}) \geq f_{\min} + \mathcal{O}\left(\frac{1}{\sqrt{n}}\right) \ .
\end{align}
As we show in Appendix~\ref{app.sdp_shannon}, the dual formulation of the Shannon entropy \gls{sdp} naturally yields such a function: its objective is affine in the observed witness $W$ and vacuum bound $\omega$, and constitutes a valid lower bound on $H(B|\Lambda_E)$ even when finite-size empirical frequencies do not admit an exact quantum realization. Explicitly, it takes the form
\begin{align}
f_{\min}(W,\omega) = \xi + \beta W + \gamma \omega \ ,
\end{align}
where $\xi$, $\beta$, and $\gamma$ are scalar coefficients obtained from the dual \gls{sdp} solution. We refer to Appendix \ref{app.eat} for details on how we estimate the non-i.i.d.~finite-size corrections.

For comparison purposes, we also benchmark our Shannon-entropy bounds with respect to the traditional approach based on bounding the min-entropy. The conditional min-entropy is given by $H_{\min}(B|\Lambda_E)=-\log_2(p_g)$, where
\begin{align}
p_g = \sum_x p_x \sum_\lambda q(\lambda) \max_b \{p_\lambda(b|x)\} \ ,
\end{align}
is the probability that the eavesdropper guesses the outcome $b$ averaged on all inputs $x$. As we detail in Appendix \ref{app.sdp_min}, the same method from Ref.~\cite{carceller2025} can be used to bound $H_{\min}(B|\Lambda_E)$ instead.

\section{Experimental Setup}\label{sec:exp_setup}

The experimental implementation was designed to combine a simple and stable transmitter with a coherent receiver characterized by low optical loss. Since the four \gls{qpsk} symbols have the same amplitude and differ only in their optical phase, they can be prepared using a single phase modulator. This avoids the bias stabilization typically required in conventional \gls{qam} transmitters based on IQ modulators. At the receiver, heterodyne detection enables the simultaneous measurement of both field quadratures and, unlike homodyne detection, does not require locking the phase of the local oscillator to a selected quadrature.
The low-loss integrated implementation of the receiver is therefore important, as optical losses at this stage directly reduce the amount of certifiable randomness.

\subsection{Optical transmitter}
%%%%%%%%%%%%%%%% bFIGURE %%%%%%%%%%%%%%%%
\begin{figure*}[htbp!]
	\centering
	\includegraphics[width=1\linewidth]{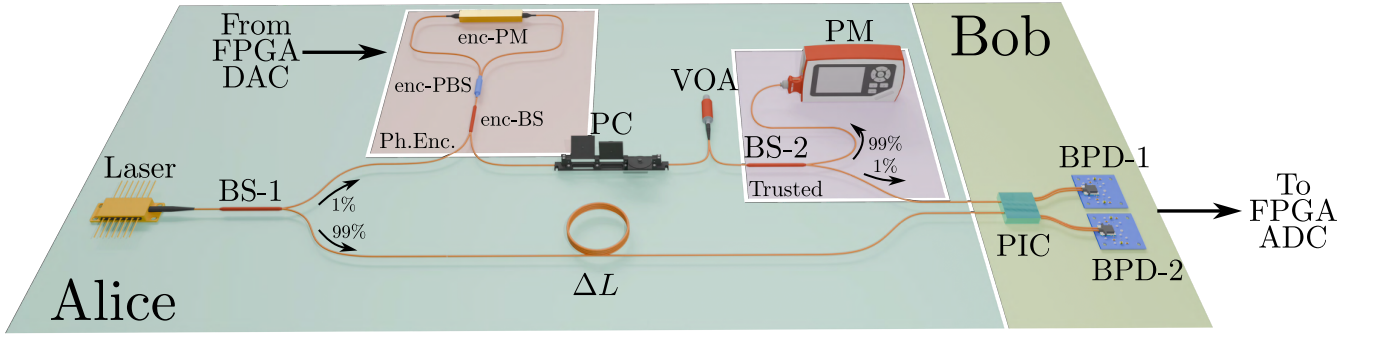}
	\caption{\textbf{Experimental QRNG scheme}.
    The figure details the schematic of the device.
    The source's light is split by a 99:1 \glsentryshort{bs} generating both the \glsentryshort{lo} and the \glsentryshort{qpsk}  quantum signal $\ket{\alpha}$ after passing through a phase encoder.
    These are then coupled at the inputs of an optical heterodyne receiver and converted into electrical signals by a pair \glspl{bpd} measuring both orthogonal quadratures of the quantum signal.
    \glsxtrshort{pc}: \glsxtrlong{pc}, \glsxtrshort{pbs}: \glsxtrlong{pbs}, \glsxtrshort{bs}: \glsxtrlong{bs}, \glsxtrshort{pm}: \glsxtrlong{pm}, \glsxtrshort{fpga}: \glsxtrlong{fpga}, \glsxtrshort{voa}: \glsxtrlong{voa}, \glsxtrshort{bpd}: \glsxtrlong{bpd}, $\Delta L$: optical delay line, \glsxtrshort{pic}: \glsxtrlong{pic}.}
    \label{fig:optical_setup}
\end{figure*}
%%%%%%%%%%%%%%%% FIGURE %%%%%%%%%%%%%%%%

A schematic representation of the experimental setup that implements our \gls{qrng} protocol is represented in Fig. (\ref{fig:optical_setup}).
The output of a \SI{1550}{\nano\meter} laser (Pure Photonics PPCL550) is split into two paths by a 99:1 \gls{bs}.
While the 1\% branch is modulated to generate a \SI{250}{\mega\baud}, \gls{qpsk} quantum signal, the remaining 99\% is used as the \gls{lo} for heterodyne detection.
This latter branch also includes an optical delay line $\Delta L$ to match the longer quantum signal path length, reducing relative phase fluctuations.
The state encoder is a simple phase modulator design, built entirely with standard fiber \gls{cots} components~\cite{Sabatini2025}.
It is based on a stable Sagnac-loop scheme, incorporating a lithium niobate phase modulator (iXblue MPZ-LN-10) driven by a \gls{fpga} (AMD Xilinx RFSoC 4$\times$2) working as a \SI{2}{\giga\hertz} arbitrary waveform generator.
After the encoder, a \gls{voa} attenuates the optical power to tune the mean photon number of the quantum signals, while a \gls{pc} aligns its polarization with that of the \gls{lo}, maximizing interference.
A second 99:1 \gls{bs} (BS-2) taps $99\%$ of light, which is measured by a \gls{pm}, to estimate the mean photon number of quantum signals, while the remaining 1\% is sent to the heterodyne receiver input.
Because this estimation is required to experimentally validate our protocol's energy-bound given in Eq.~\ref{eq:sdi_ass}, the power meter PM must be calibrated and the splitting ratio of BS-2 must be accurately known, no other component in the setup requires precise characterization.

\subsection{Optical receiver}
The heterodyne receiver is implemented using a glass-based integrated tunable 90-degree hybrid, whose outputs are connected to a pair of \SI{2.5}{\giga\hertz} balanced photodetectors (Thorlabs PDB780CAC).
The resulting electrical signals are then filtered by a 500 \SI{}{\mega\hertz} low-pass filter, and digitized by an \gls{fpga} (AMD Xilinx RFSoC 4$\times$2) operating at a sampling rate of 2 \SI{}{\giga\sample/\second} with 14-bit resolution.
Finally, the digitized data is streamed to a PC for further off-line processing, consisting of \gls{dsp} and randomness certification.

The \gls{pic} at the heart of our heterodyne detector, whose design, fabrication, and characterization are detailed in~\cite{Peri2025}, features a polarization-independent, low-loss architecture manufactured with \gls{flm}~\cite{Corrielli2021}.
This technique focuses short, intense laser pulses into a transparent substrate to directly inscribe waveguides in the material.
This technique has a broad range of capabilities useful for optical information processing, such as polarization insensitivity and low insertion losses.
The detection device, which presents insertion losses lower than $\SI{1.28}{\decibel}$, is finely tuned to work as a heterodyne receiver.
This configuration requires the maximization of the \gls{cmrr} and optimization of the phase shift between the two simultaneous measurements of the input state's quadratures to be orthogonal.
From our estimations, we obtained a high \gls{cmrr} value greater than \SI{73}{\decibel} and a precise $\SI{90}{\degree}$ phase shift with a phase error as low as $\SI{0.9}{\degree}$.
The receiver also exhibits a stable behaviour, maintaining its optimal operating point for several hours without the need for an external feedback control loop apart from temperature stabilization~\cite{Peri2025}.
Furthermore, it is worth noticing that although our \gls{qrng} protocol can extract randomness regardless of the receiver configuration, the ability to accurately set and preserve its optimal configuration is essential to guarantee the maximum amount of entropy extractable from the experimental setup.

\section{Results}\label{sec:results}
\begin{figure}[htbp!]
    \centering
    \includegraphics[scale=0.58]{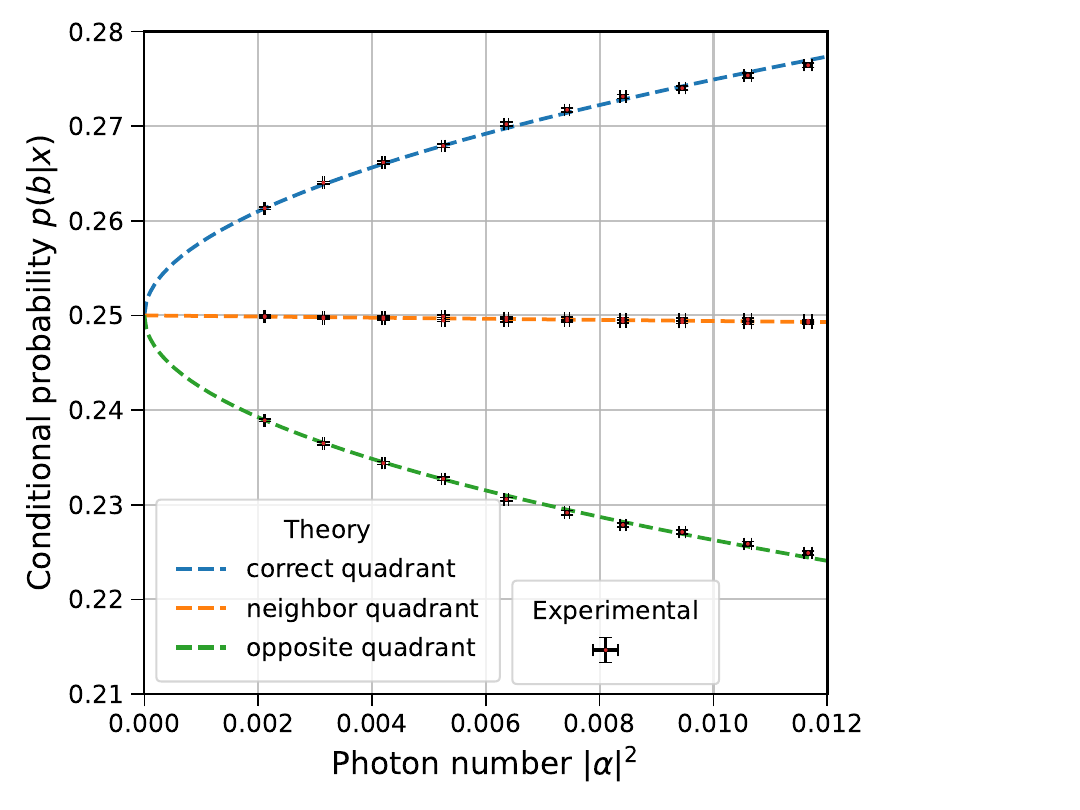}
    \caption{\textbf{Conditional probabilities estimation.} Estimated conditional probabilities $p(b|x)$ as a function of the mean photon number $|\alpha|^2$ of the \gls{qpsk}-modulated coherent states.
    For each value of $|\alpha|^2$, the probabilities are obtained by averaging over the four \gls{qpsk} input symbols and grouping the outcomes according to whether Bob's detected quadrant is the \emph{correct} quadrant (same quadrant as the prepared symbol), the \emph{opposite} quadrant, or one of the two \emph{neighboring} quadrants.
    The values are averaged over $50$ independent acquisitions, each containing approximately one million symbols.
    Error bars account for statistical and optical-power uncertainties.
    The dashed curves show the corresponding theoretical model evaluated at the best-fit detection efficiency $\eta$.}
    \label{fig:fitted_plot_eta}
\end{figure}

To quantify the amount of randomness that can be certified from our system, we first estimated the conditional outcome probabilities $p(b|x)$ for different values of the experimental mean photon number $|\alpha|^2$.
The results are presented in Fig.~\ref{fig:fitted_plot_eta}.
For each value of $|\alpha|^2$, the probabilities were obtained from $50$ independent acquisitions, each containing approximately one million symbols.
As expected, the probability associated with the correct quadrant detections increases with $|\alpha|^2$, while the probabilities of assigning the outcome to the opposite or neighboring quadrants decrease accordingly.
This behavior reflects the better distinguishability of the prepared coherent states as their separation in phase space increases.
It is also worth noting that the measured curve $\vert \alpha \vert^2$-$p(b \vert x)$ is consistent with the theoretical model for \gls{qpsk} modulation under heterodyne detection, when the overall system efficiency $\eta$ is also considered, as illustrated by the dashed curves in Fig.~\ref{fig:fitted_plot_eta}. The latter was estimated by considering a free-parameter term in the theoretical model of the device, which was then used to fit the witness (probability of correct detection), yielding an effective $\eta$ value of approximately $0.38$. The experimental data were acquired around the entropy maximum expected from the theoretical model.

%%%%%%%%%%%%%%%% bFIGURE %%%%%%%%%%%%%%%%
%%%%%%%%%%%%%%%% eFIGURE %%%%%%%%%%%%%%%%

%%%%%%%%%%%%%%%% bFIGURE %%%%%%%%%%%%%%%%
\begin{figure*}[htbp!]
    \centering
    \includegraphics[scale=0.65]{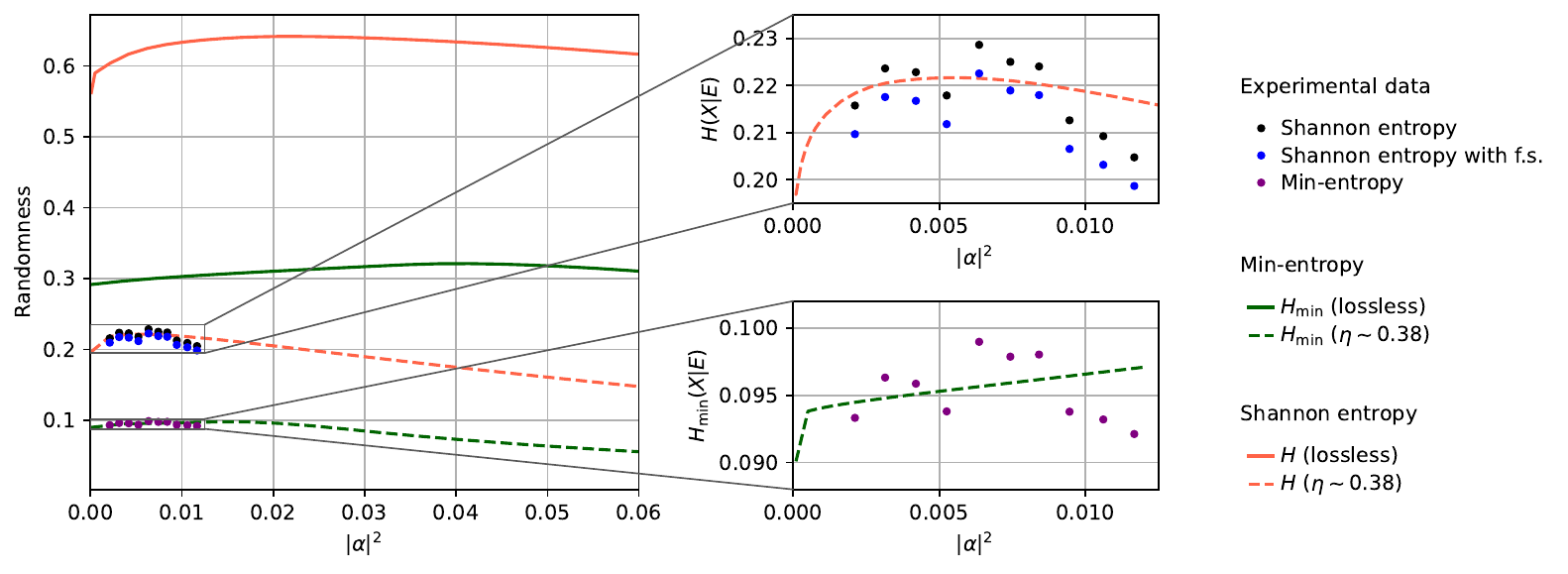}
    \caption{\textbf{Certifiable randomness.} The figure details the amount of randomness that can be experimentally certified per measurement: the Shannon entropy values are shown in blue, their finite‑size correction obtained via the EAT framework in purple, and the min‑entropy in black.
    The figure also includes the theoretical performance of the \gls{qrng}, with solid curves indicating the ideal case (no losses) and dashed curves representing the predictions when losses are considered.
    }
    \label{fig:randomness}
\end{figure*}
%%%%%%%%%%%%%%%% eFIGURE %%%%%%%%%%%%%%%%

Based on the measured conditional probabilities, we then evaluated the certifiable randomness generated by the protocol. The results are reported in Fig.~\ref{fig:randomness}.
The maximum certifiable conditional Shannon entropy per measurement is approximately \SI{0.229}{bit} per measurement and is obtained for $|\alpha|^2 \sim 0.006$.
When including finite-size effects through \gls{eat} in a non-i.i.d. scenario, the per-symbol conditional smooth min-entropy $H_{\min}^\varepsilon(\mathbf{B}|\mathbf{E})$ decreases slightly due to finite-size effects, yielding approximately \SI{0.223}{bit} per measurement.
With a symbol rate of \SI{250}{MBaud}, this corresponds to a secure generation rate of \SI{55.6}{\mega bit/\second}.
For comparison, Fig.~\ref{fig:randomness} also reports the more conservative $H_{\min}^\varepsilon(\mathbf{B}|\mathbf{E})$ lower bound based on the conditional min-entropy $H_{\rm min} (\mathbf{B}|\mathbf{E})$, which reaches approximately \SI{0.099}{bit} per measurement for the same value of $|\alpha|^2$, leading to \SI{24.7}{\mega bit/\second} of generation rate, with a maximum predicted for a slightly higher $\vert \alpha \vert^2$ of $\sim 0.014$.
This comparison clearly highlights the advantage of directly certifying the Shannon entropy, rather than relying solely on min-entropy bounds.
Notably, the experimental rates we estimated follow the theoretical predictions once the system's efficiency $\eta\sim 0.38$ is accounted for, as illustrated in Fig.~\ref{fig:randomness} with dashed curves.
For completeness, we also reported the achievable rates in the ideal case of unit efficiency.

Our results are summarized and compared with other experimental implementations of \gls{sdi}-\gls{qrng} protocols in Tab.~\ref{tab:semiDI-QRNG}.
\begin{table*}[htbp!]
\small
\renewcommand{\arraystretch}{1.15}
\setlength{\tabcolsep}{5pt}
\centering
\resizebox{\textwidth}{!}{%
\begin{tabular}{ccccccccc}
\hline
\multirow{2}{*}{Reference} &
\multirow{2}{*}{Bound} &
\makecell[c]{non-i.i.d.\\assumption\\} &
\makecell[c]{Implementation \\ (Discrete/Continuous \\ Variables)} &
\makecell[c]{State\\preparation\\} &
\makecell[c]{$H_{\min}^\varepsilon(\mathbf{B}|\mathbf{E})$ from\\min-entropy bound\\ (bit/meas)} &
\makecell[c]{$H_{\min}^\varepsilon(\mathbf{B}|\mathbf{E})$ from\\Shannon bound\\ (bit/meas)} &
\makecell[c]{Repetition\\rate [$\si{\mega\hertz}$]\\} &
\makecell[c]{Randomness \\ generation\\rate [$\mathrm{Mbit/s}$]\\} \\
&&&&&&&&\\[-0.8em]

\hline

\cite{Lunghi2015} & dimension & $\times$ & DV & polarization & -- & -- & $0.03$ & $23\cdot 10^{-6}$ \\

\cite{Brask2017} & overlap & $\times$ & DV & time-bin & $0.22$ & -- & $50$ & $11$ \\

\cite{Brask2017} & overlap & $\times$ & DV & on-off keying & $0.33$ & -- & $50$ & $16.5$ \\

\cite{Leone2020} & overlap & $\times$ & DV & on-off keying & $0.0061$ & --  & $1$ & $0.006$ \\

\cite{carceller2025improving} & overlap & $\checkmark$ & DV & time-bin & -- & $1.299^{\ddagger}$ & $0.526$ & $0.683^{*}$ \\

\cite{Tebyanian2021_2nd} & energy & $\times$ & DV & time-bin & $0.23$ & -- & $10$ & $2.3^*$ \\

\cite{rusca_self-testing_2019} & energy & $\checkmark$ & DV & on-off keying & -- & $0.1$ & $12.5$ & $1.25$ \\

\cite{Avesani2021} & energy & $\times$ & CV & BPSK & $0.09$ & -- & $1250$ & $113$ \\

\cite{lu_semi-device-independent_2026} & energy & $\times$ & CV & ternary coherent states & $0.0117$  & -- & $100$ & $1.165$ \\

\cite{rusca_fast_2020} & energy & $\checkmark$ & CV & BPSK & -- & $0.12\,(0.18^{*})$ & $1250$ & $145.5$ \\

This work & energy & $\checkmark$ & CV & QPSK & 0.10 & $0.22^{\ddagger}$ & $250$ & $55.6$ \\
\hline
\end{tabular}
}%
\caption{Comparative analysis of Semi-DI QRNG implementations with untrusted source and measurement. The comparison includes the underlying protocol assumptions, such as the bound on the quantum states and the requirement of an independent and identically distributed (i.i.d.) assumption, together with relevant implementation parameters and certified performance. Values marked with $*$ were not explicitly reported in the corresponding reference and were estimated from the available data. For Ref.~\cite{rusca_fast_2020}, $0.12\,(0.18^*)$ denotes the entropy threshold used to certify the reported generation rate, with the value in parentheses indicating the maximum entropy per measurement inferred from the reported data. Values marked with $^\ddagger$ were obtained using the entropy accumulation theorem (EAT).}
\label{tab:semiDI-QRNG}
\end{table*}
In general, \gls{dv} protocols achieve the highest certifiable randomness per round~\cite{Brask2017, Tebyanian2021_2nd}, enabled by the high detection efficiency and low noise of single‑photon detectors. By contrast, coherent‑receiver–based \gls{cv} implementations typically face lower detection efficiency and higher noise, although steady progress continues to be made at the industrial level.
Despite this, \gls{cv} systems are often easier to implement and generally allow for significantly higher repetition rates than \gls{dv}, compensating for the lower per-round entropy certification. 
Although Refs.~\cite{Avesani2021,rusca_fast_2020} report higher speeds, these values are due to their larger repetition rates. In contrast, among the reported \gls{cv} implementations, our protocol achieves the highest certifiable randomness per measurement to date. Our limit is mainly caused by the bandwidth of the generation and acquisition electronic chain available.  Nevertheless, high-order discrete-modulated coherent-state implementations have already demonstrated operation at \SI{1}{GBaud}~\cite{pan_experimental_2022}, and integrated coherent CV systems have reached \SI{10}{GBaud} operation~\cite{Hajomer2024}. Assuming the same certified entropy per measurement,  our protocol could achieve rates of approximately \SI{223}{Mbit/s} and \SI{2.23}{Gbit/s}, respectively.
The high certifiable entropy per measurement achieved in our experiment is enabled by three key improvements: a randomness certification approach that yields tighter bounds on the conditional smooth min-entropy $H_{\min}^\varepsilon(\mathbf{B}|\mathbf{E})$, the adoption of the more complex \gls{qpsk} modulation format, and the use of a low-loss heterodyne receiver.
Concerning the latter, optical losses at the receiver stage in practical implementations can significantly reduce the amount of certifiable randomness.
The low-loss performance of our heterodyne receiver is therefore crucial to maximize the efficiency of the setup, and consequently, the randomness generation rate.
Conventional $\SI{90}{\degree}$ optical hybrid receivers realized on silicon platforms typically exhibit optical efficiencies between $0.31$ and $0.56$, excluding detector efficiency \cite{Raffaelli2018, Zhang2019, Hajomer2024, Bian2024}. By contrast, our glass-based integrated photonic receiver achieves an optical efficiency of $\eta_{\mathrm{PIC}} \sim 0.79$.
However, as previously reported, the overall efficiency of the system was estimated to be approximately $0.38$. In addition to optical losses and the quantum efficiency of the external photodetectors ($\eta_{\mathrm{BPD}} \sim 0.77$), this values also accounts for other effective losses, including electronic noise, residual mode mismatch, and imperfect optical alignment of the setup (with the polarization alignment being manually optimized).

\section{Conclusions}\label{sec:Conclusion}

In this work, we present a simple and practical \gls{sdi}-\gls{qrng} based, for the first time, on an energy-constrained framework supporting \gls{qpsk} coherent-state modulation in the finite-size non-i.i.d.~regime. The protocol leverages an assumption on the average vacuum component of the prepared quantum states, without requiring a detailed characterization of either the source or the receiver. Within this framework, we directly lower-bound the single-shot conditional Shannon entropy of the measurement outcomes through a hierarchy of \glspl{sdp} and use it as a min-tradeoff function within the \gls{eat} framework to certify randomness in the finite-size non-i.i.d.~regime.
Our experimental implementation combines a practical fiber-based state encoder on the transmitter side with a stable, low-loss integrated heterodyne receiver, providing a compact and telecom-compatible architecture for high-rate \gls{sdi} randomness generation.
Building on this architecture, our \gls{sdi}-\gls{qrng} achieves a lower bound of \SI{0.223}{bit} per measurement on the conditional smooth min-entropy with \gls{eat}, representing, to the best of our knowledge, the highest level of entropy certification achieved to date for \gls{cv} systems.
This result highlights the relevance of both the theoretical framework underlying the employed randomness certification and the low-loss architecture of our heterodyne receiver.
We believe that our \gls{sdi}-\gls{qrng} represents an interesting solution for secure, robust, and high-rate randomness generation, readily integrable into existing communication infrastructures thanks to the use of fiber-based components and a rapidly manufacturable \gls{pic}.

Further improvements of the platform could be achieved through the full integration of both the transmitter and receiver within photonic integrated circuits. Such an approach would enable higher repetition rates, improved stability, and a significant reduction in system complexity, ultimately facilitating deployment in real-world, out-of-the-laboratory scenarios. \\

\acknowledgments
This work was partially supported by European Union’s Horizon Europe research and innovation program under the project Quantum Secure Networks Partnership (QSNP), grant agreement No 101114043. Author M.R.B. acknowledges support from the European Union’s Horizon Europe Framework Programme under the Marie Sklodowska Curie Grant No. 101072637, Project Quantum-Safe Internet (QSI). Views and opinions expressed are however those of the authors only and do not necessarily reflect those of the European Union or European Commission-EU. Neither the European Union nor the granting authority can be held responsible for them. C.R.C. and A.T. are financially supported by the Swedish Foundation for Strategic Research, the Knut and Alice Wallenberg Foundation through the Wallenberg Center for Quantum Technology (WACQT) and the Swedish Research Council under Contract No. 2023-03498. 

\section*{Code availability}

The codes used to generate the results in this paper are available in GitHub: \url{https://github.com/chalswater//QRNG_with_Heterodyne}.

\bibliography{main_bib.bib}

@article{rusca_self-testing_2019,
	title = {Self-testing quantum random-number generator based on an energy bound},
	volume = {100},
	url = {https://link.aps.org/doi/10.1103/PhysRevA.100.062338},
	doi = {10.1103/PhysRevA.100.062338},
	number = {6},
	urldate = {2025-12-11},
	journal = {Physical Review A},
	author = {Rusca, Davide and van Himbeeck, Thomas and Martin, Anthony and Brask, Jonatan Bohr and Shi, Weixu and Pironio, Stefano and Brunner, Nicolas and Zbinden, Hugo},
	month = dec,
	year = {2019},
	note = {Publisher: American Physical Society},
	pages = {062338},
}

@misc{Brask2026,
      title={Quantum correlations in prepare-and-measure scenarios and their semi-device-independent applications}, 
      author={Jonatan Bohr Brask and Nicolas Brunner and Jef Pauwels and Davide Rusca and Armin Tavakoli},
      year={2026},
      eprint={2603.23604},
      archivePrefix={arXiv},
      primaryClass={quant-ph},
      url={https://arxiv.org/abs/2603.23604}, 
}

@article{Pauwels2022,
  title = {Almost Qudits in the Prepare-and-Measure Scenario},
  author = {Pauwels, Jef and Pironio, Stefano and Woodhead, Erik and Tavakoli, Armin},
  journal = {Phys. Rev. Lett.},
  volume = {129},
  issue = {25},
  pages = {250504},
  numpages = {7},
  year = {2022},
  month = {Dec},
  publisher = {American Physical Society},
  doi = {10.1103/PhysRevLett.129.250504},
  url = {https://link.aps.org/doi/10.1103/PhysRevLett.129.250504}
}

@article{SDPreview,
  title = {Semidefinite programming relaxations for quantum correlations},
  author = {Tavakoli, Armin and Pozas-Kerstjens, Alejandro and Brown, Peter and Ara\'ujo, Mateus},
  journal = {Rev. Mod. Phys.},
  volume = {96},
  issue = {4},
  pages = {045006},
  numpages = {68},
  year = {2024},
  month = {Dec},
  publisher = {American Physical Society},
  doi = {10.1103/RevModPhys.96.045006},
  url = {https://link.aps.org/doi/10.1103/RevModPhys.96.045006}
}

@article{rusca_fast_2020,
	title = {Fast self-testing quantum random number generator based on homodyne detection},
	volume = {116},
	issn = {0003-6951},
	url = {https://doi.org/10.1063/5.0011479},
	doi = {10.1063/5.0011479},
	number = {26},
	urldate = {2025-12-11},
	journal = {Applied Physics Letters},
	author = {Rusca, Davide and Tebyanian, Hamid and Martin, Anthony and Zbinden, Hugo},
	month = jul,
	year = {2020},
	pages = {264004},
}

@article{Sabatini2025,
    author = {Sabatini, Mattia and Bertapelle, Tommaso and Villoresi, Paolo and Vallone, Giuseppe and Avesani, Marco},
    title = {Hybrid Encoder for Discrete and Continuous Variable QKD},
    journal = {Advanced Quantum Technologies},
    volume = {8},
    number = {8},
    pages = {2400522},
    doi = {https://doi.org/10.1002/qute.202400522},
    year = {2025}
}

@article{Peri2025,
   title={High-performance heterodyne receiver for quantum information processing in a laser-written integrated photonic platform},
   volume={8},
   ISSN={2577-5421},
   url={http://dx.doi.org/10.1117/1.AP.8.1.016009},
   DOI={10.1117/1.ap.8.1.016009},
   number={01},
   journal={Advanced Photonics},
   publisher={SPIE-Intl Soc Optical Eng},
   author={Peri, Andrea and Gualandi, Giulio and Bertapelle, Tommaso and Sabatini, Mattia and Corrielli, Giacomo and Piétri, Yoann and Marangon, Davide Giacomo and Vallone, Giuseppe and Villoresi, Paolo and Osellame, Roberto and Avesani, Marco},
   year={2026},
   month=Feb }

@article{carceller2025,
  title = {Prepare-and-Measure Scenarios with Photon-Number Constraints},
  author = {Roch i Carceller, Carles and Pauwels, Jef and Pironio, Stefano and Tavakoli, Armin},
  journal = {Phys. Rev. Lett.},
  volume = {135},
  issue = {14},
  pages = {140802},
  numpages = {8},
  year = {2025},
  month = {Oct},
  publisher = {American Physical Society},
  doi = {10.1103/glty-kkbp},
  url = {https://link.aps.org/doi/10.1103/glty-kkbp}
}

@Article{Avesani2018,
    author={Avesani, Marco
    and Marangon, Davide G.
    and Vallone, Giuseppe
    and Villoresi, Paolo},
    title={Source-device-independent heterodyne-based quantum random number generator at 17 Gbps},
    journal={Nature Communications},
    year={2018},
    month={Dec},
    day={18},
    volume={9},
    number={1},
    pages={5365},
}

@article{Pan2015,
    author = {Nie, You-Qi and Huang, Leilei and Liu, Yang and Payne, Frank and Zhang, Jun and Pan, Jian-Wei},
    title = {The generation of 68 Gbps quantum random number by measuring laser phase fluctuations},
    journal = {Review of Scientific Instruments},
    volume = {86},
    number = {6},
    pages = {063105},
    year = {2015},
    month = {06},
    issn = {0034-6748},
    doi = {10.1063/1.4922417},
    url = {https://doi.org/10.1063/1.4922417},
}

@article{Bruynsteen2023,
    title = {100-Gbit/s Integrated Quantum Random Number Generator Based on Vacuum Fluctuations},
    author = {Bruynsteen, C\'edric and Gehring, Tobias and Lupo, Cosmo and Bauwelinck, Johan and Yin, Xin},
    journal = {PRX Quantum},
    volume = {4},
    issue = {1},
    pages = {010330},
    numpages = {11},
    year = {2023},
    month = {Mar},
    publisher = {American Physical Society},
    doi = {10.1103/PRXQuantum.4.010330},
    url = {https://link.aps.org/doi/10.1103/PRXQuantum.4.010330}
}

@article{Pironio2010,
    author={Pironio, S.
    and Ac{\'i}n, A.
    and Massar, S.
    and de la Giroday, A. Boyer
    and Matsukevich, D. N.
    and Maunz, P.
    and Olmschenk, S.
    and Hayes, D.
    and Luo, L.
    and Manning, T. A.
    and Monroe, C.},
    title={Random numbers certified by Bell's theorem},
    journal={Nature},
    year={2010},
    month={Apr},
    day={01},
    volume={464},
    number={7291},
    pages={1021-1024},
    doi={10.1038/nature09008},
    url={https://doi.org/10.1038/nature09008}
}

@article{Cao2015,
    doi = {10.1088/1367-2630/17/12/125011},
    url = {https://doi.org/10.1088/1367-2630/17/12/125011},
    year = {2015},
    month = {dec},
    publisher = {IOP Publishing},
    volume = {17},
    number = {12},
    pages = {125011},
    author = {Cao, Zhu and Zhou, Hongyi and Ma, Xiongfeng},
    title = {Loss-tolerant measurement-device-independent quantum random number generation},
    journal = {New Journal of Physics},
}

@article{Marangon2019,
    title = {Simple source device-independent continuous-variable quantum random number generator},
    author = {Smith, P. R. and Marangon, D. G. and Lucamarini, M. and Yuan, Z. L. and Shields, A. J.},
    journal = {Phys. Rev. A},
    volume = {99},
    issue = {6},
    pages = {062326},
    numpages = {8},
    year = {2019},
    month = {Jun},
    publisher = {American Physical Society},
    doi = {10.1103/PhysRevA.99.062326},
    url = {https://link.aps.org/doi/10.1103/PhysRevA.99.062326}
}

@article{Lunghi2015,
    title = {Self-Testing Quantum Random Number Generator},
    author = {Lunghi, Tommaso and Brask, Jonatan Bohr and Lim, Charles Ci Wen and Lavigne, Quentin and Bowles, Joseph and Martin, Anthony and Zbinden, Hugo and Brunner, Nicolas},
    journal = {Phys. Rev. Lett.},
    volume = {114},
    issue = {15},
    pages = {150501},
    numpages = {5},
    year = {2015},
    month = {Apr},
    publisher = {American Physical Society},
    doi = {10.1103/PhysRevLett.114.150501},
    url = {https://link.aps.org/doi/10.1103/PhysRevLett.114.150501}
}

@article{VanHimbeeck2017,
    doi = {10.22331/q-2017-11-18-33},
    url = {https://doi.org/10.22331/q-2017-11-18-33},
    title = {Semi-device-independent framework based on natural physical assumptions},
    author = {Van Himbeeck, Thomas and Woodhead, Erik and Cerf, Nicolas J. and Garc{\'{i}}a-Patr{\'{o}}n, Ra{\'{u}}l and Pironio, Stefano},
    journal = {{Quantum}},
    issn = {2521-327X},
    publisher = {{Verein zur F{\"{o}}rderung des Open Access Publizierens in den Quantenwissenschaften}},
    volume = {1},
    pages = {33},
    month = nov,
    year = {2017}
}

@article{Avesani2021,
    title = {Semi-Device-Independent Heterodyne-Based Quantum Random-Number Generator},
    author = {Avesani, Marco and Tebyanian, Hamid and Villoresi, Paolo and Vallone, Giuseppe},
    journal = {Phys. Rev. Appl.},
    volume = {15},
    issue = {3},
    pages = {034034},
    numpages = {11},
    year = {2021},
    month = {Mar},
    publisher = {American Physical Society},
    doi = {10.1103/PhysRevApplied.15.034034},
    url = {https://link.aps.org/doi/10.1103/PhysRevApplied.15.034034}
}

@article{Tebyanian2021_2nd,
    doi = {10.1088/2058-9565/ac2047},
    url = {https://doi.org/10.1088/2058-9565/ac2047},
    year = {2021},
    month = {sep},
    publisher = {IOP Publishing},
    volume = {6},
    number = {4},
    pages = {045026},
    author = {Tebyanian, Hamid and Zahidy, Mujtaba and Avesani, Marco and Stanco, Andrea and Villoresi, Paolo and Vallone, Giuseppe},
    title = {Semi-device independent randomness generation based on quantum state’s indistinguishability},
    journal = {Quantum Science and Technology},
}

@article{Brask2017,
    title = {Megahertz-Rate Semi-Device-Independent Quantum Random Number Generators Based on Unambiguous State Discrimination},
    author = {Brask, Jonatan Bohr and Martin, Anthony and Esposito, William and Houlmann, Raphael and Bowles, Joseph and Zbinden, Hugo and Brunner, Nicolas},
    journal = {Phys. Rev. Appl.},
    volume = {7},
    issue = {5},
    pages = {054018},
    numpages = {10},
    year = {2017},
    month = {May},
    publisher = {American Physical Society},
    doi = {10.1103/PhysRevApplied.7.054018},
    url = {https://link.aps.org/doi/10.1103/PhysRevApplied.7.054018}
}

@article{Leone2020,
    author = {Leone, Nicolò and Rusca, Davide and Azzini, Stefano and Fontana, Giorgio and Acerbi, Fabio and Gola, Alberto and Tontini, Alessandro and Massari, Nicola and Zbinden, Hugo and Pavesi, Lorenzo},
    title = {An optical chip for self-testing quantum random number generation},
    journal = {APL Photonics},
    volume = {5},
    number = {10},
    pages = {101301},
    year = {2020},
    month = {10},
    issn = {2378-0967},
    doi = {10.1063/5.0022526},
    url = {https://doi.org/10.1063/5.0022526},
}

@article{Corrielli2021,
    url = {https://doi.org/10.1515/nanoph-2021-0419},
    title = {Femtosecond laser micromachining for integrated quantum photonics},
    author = {Giacomo Corrielli and Andrea Crespi and Roberto Osellame},
    pages = {3789--3812},
    volume = {10},
    number = {15},
    journal = {Nanophotonics},
    doi = {doi:10.1515/nanoph-2021-0419},
    year = {2021},
    lastchecked = {2024-12-17}
}

@article{Raffaelli2018,
  title={A homodyne detector integrated onto a photonic chip for measuring quantum states and generating random numbers},
  author={Raffaelli, Francesco and Ferranti, Giacomo and Mahler, Dylan H and Sibson, Philip and Kennard, Jake E and Santamato, Alberto and Sinclair, Gary and Bonneau, Damien and Thompson, Mark G and Matthews, Jonathan CF},
  journal={Quantum Science and Technology},
  volume={3},
  number={2},
  pages={025003},
  year={2018},
  publisher={IOP Publishing}
}

@Article{Zhang2019,
    author={Zhang, G.
    and Haw, J. Y.
    and Cai, H.
    and Xu, F.
    and Assad, S. M.
    and Fitzsimons, J. F.
    and Zhou, X.
    and Zhang, Y.
    and Yu, S.
    and Wu, J.
    and Ser, W.
    and Kwek, L. C.
    and Liu, A. Q.},
    title={An integrated silicon photonic chip platform for continuous-variable quantum key distribution},
    journal={Nature Photonics},
    year={2019},
    month={Dec},
    day={01},
    volume={13},
    number={12},
    pages={839-842},
    doi={10.1038/s41566-019-0504-5},
    url={https://doi.org/10.1038/s41566-019-0504-5}
}

@article{Hajomer2024,
    title={Continuous-variable quantum key distribution at 10 gbaud using an integrated photonic-electronic receiver},
    author={Hajomer, Adnan AE and Bruynsteen, C{\'e}dric and Derkach, Ivan and Jain, Nitin and Bomhals, Axl and Bastiaens, Sarah and Andersen, Ulrik L and Yin, Xin and Gehring, Tobias},
    journal={Optica},
    volume={11},
    number={9},
    pages={1197--1204},
    year={2024},
    publisher={Optica Publishing Group}
}

@article{Bian2024,
    author = {Bian, Yiming and Pan, Yan and Xu, Xuesong and Zhao, Liang and Li, Yang and Huang, Wei and Zhang, Lei and Yu, Song and Zhang, Yichen and Xu, Bingjie},
    title = {Continuous-variable quantum key distribution over 28.6 km fiber with an integrated silicon photonic receiver chip},
    journal = {Applied Physics Letters},
    volume = {124},
    number = {17},
    pages = {174001},
    year = {2024},
    month = {04},
    issn = {0003-6951},
    doi = {10.1063/5.0203130}
}

@INPROCEEDINGS{metger2022,
    author={Metger, Tony and Fawzi, Omar and Sutter, David and Renner, Renato},
    booktitle={2022 IEEE 63rd Annual Symposium on Foundations of Computer Science (FOCS)}, 
    title={Generalised entropy accumulation}, 
    year={2022},
    volume={},
    number={},
    pages={844-850},
    doi={10.1109/FOCS54457.2022.00085}
}

@article{brown2024,
    doi = {10.22331/q-2024-08-27-1445},
    url = {https://doi.org/10.22331/q-2024-08-27-1445},
    title = {Device-independent lower bounds on the conditional von {N}eumann entropy},
    author = {Brown, Peter and Fawzi, Hamza and Fawzi, Omar},
    journal = {{Quantum}},
    issn = {2521-327X},
    publisher = {{Verein zur F{\"{o}}rderung des Open Access Publizierens in den Quantenwissenschaften}},
    volume = {8},
    pages = {1445},
    month = aug,
    year = {2024}
}

@article{carceller2025improving,
    title = {Improving semi-device-independent randomness certification by entropy accumulation},
    author = {Carceller, Carles Roch i and Faria, Lucas Nunes and Liu, Zheng-Hao and Sguerso, Nicol\`o and Andersen, Ulrik Lund and Neergaard-Nielsen, Jonas Schou and Brask, Jonatan Bohr},
    journal = {Phys. Rev. A},
    volume = {112},
    issue = {2},
    pages = {022430},
    numpages = {10},
    year = {2025},
    month = {Aug},
    publisher = {American Physical Society},
    doi = {10.1103/dwdv-89bj},
    url = {https://link.aps.org/doi/10.1103/dwdv-89bj}
}

@article{Bancal2014,
    doi = {10.1088/1367-2630/16/3/033011},
    url = {https://doi.org/10.1088/1367-2630/16/3/033011},
    year = {2014},
    month = {mar},
    publisher = {IOP Publishing},
    volume = {16},
    number = {3},
    pages = {033011},
    author = {Bancal, Jean-Daniel and Sheridan, Lana and Scarani, Valerio},
    title = {More randomness from the same data},
    journal = {New Journal of Physics}
}

@article{Acin2016,
	author = {Ac{\'\i}n, Antonio and Masanes, Lluis},
	date = {2016/12/01},
	doi = {10.1038/nature20119},
	id = {Ac{\'\i}n2016},
	isbn = {1476-4687},
	journal = {Nature},
	number = {7632},
	pages = {213--219},
	title = {Certified randomness in quantum physics},
	url = {https://doi.org/10.1038/nature20119},
	volume = {540},
	year = {2016}}

@article{Li2012,
  title = {Semi-device-independent randomness certification using $n\ensuremath{\rightarrow}1$ quantum random access codes},
  author = {Li, Hong-Wei and Paw\l{}owski, Marcin and Yin, Zhen-Qiang and Guo, Guang-Can and Han, Zheng-Fu},
  journal = {Phys. Rev. A},
  volume = {85},
  issue = {5},
  pages = {052308},
  numpages = {4},
  year = {2012},
  month = {May},
  publisher = {American Physical Society},
  doi = {10.1103/PhysRevA.85.052308},
  url = {https://link.aps.org/doi/10.1103/PhysRevA.85.052308}
}

@article{Li2011,
  title = {Semi-device-independent random-number expansion without entanglement},
  author = {Li, Hong-Wei and Yin, Zhen-Qiang and Wu, Yu-Chun and Zou, Xu-Bo and Wang, Shuang and Chen, Wei and Guo, Guang-Can and Han, Zheng-Fu},
  journal = {Phys. Rev. A},
  volume = {84},
  issue = {3},
  pages = {034301},
  numpages = {4},
  year = {2011},
  month = {Sep},
  publisher = {American Physical Society},
  doi = {10.1103/PhysRevA.84.034301},
  url = {https://link.aps.org/doi/10.1103/PhysRevA.84.034301}
}

@misc{vanhimbeeck2019,
      title={Correlations and randomness generation based on energy constraints}, 
      author={Thomas Van Himbeeck and Stefano Pironio},
      year={2019},
      eprint={1905.09117},
      archivePrefix={arXiv},
      primaryClass={quant-ph},
      url={https://arxiv.org/abs/1905.09117}, 
}

@misc{lu_semi-device-independent_2026,
	title = {Semi-{Device}-{Independent} {Quantum} {Random} {Number} {Generator} {Resistant} to {General} {Attacks}},
	url = {http://arxiv.org/abs/2602.06362},
	doi = {10.48550/arXiv.2602.06362},
	urldate = {2026-04-01},
	publisher = {arXiv},
	author = {Lu, Zhenguo and Wu, Jundong and Zhang, Yu and Ren, Shaobo and Wang, Xuyang and Zhou, Hongyi and Li, Yongmin},
	month = feb,
	year = {2026},
	note = {arXiv:2602.06362 [quant-ph]},
}

@phdthesis{Renner05,
    author       = {Renato Renner},
    title        = {Security of Quantum Key Distribution},
    year         = {2005},
    month        = {9},
    note         = {Available at http://arxiv.org/abs/quant-ph/0512258},
    school       = {{ETH Zurich}},
}

@article{bierhorst_experimentally_2018,
	title = {Experimentally generated randomness certified by the impossibility of superluminal signals},
	volume = {556},
	copyright = {2018 Macmillan Publishers Ltd., part of Springer Nature},
	issn = {1476-4687},
	url = {https://www.nature.com/articles/s41586-018-0019-0},
	doi = {10.1038/s41586-018-0019-0},
	number = {7700},
	urldate = {2026-05-17},
	journal = {Nature},
	publisher = {Nature Publishing Group},
	author = {Bierhorst, Peter and Knill, Emanuel and Glancy, Scott and Zhang, Yanbao and Mink, Alan and Jordan, Stephen and Rommal, Andrea and Liu, Yi-Kai and Christensen, Bradley and Nam, Sae Woo and Stevens, Martin J. and Shalm, Lynden K.},
	month = apr,
	year = {2018},
	pages = {223--226},
}

@article{li_experimental_2021,
	title = {Experimental {Realization} of {Device}-{Independent} {Quantum} {Randomness} {Expansion}},
	volume = {126},
	url = {https://link.aps.org/doi/10.1103/PhysRevLett.126.050503},
	doi = {10.1103/PhysRevLett.126.050503},
	number = {5},
	urldate = {2026-05-17},
	journal = {Physical Review Letters},
	publisher = {American Physical Society},
	author = {Li, Ming-Han and Zhang, Xingjian and Liu, Wen-Zhao and Zhao, Si-Ran and Bai, Bing and Liu, Yang and Zhao, Qi and Peng, Yuxiang and Zhang, Jun and Zhang, Yanbao and Munro, W. J. and Ma, Xiongfeng and Zhang, Qiang and Fan, Jingyun and Pan, Jian-Wei},
	month = feb,
	year = {2021},
	pages = {050503},
}

@article{liu_device-independent_2018,
	title = {Device-independent quantum random-number generation},
	volume = {562},
	copyright = {2018 Springer Nature Limited},
	issn = {1476-4687},
	url = {https://www.nature.com/articles/s41586-018-0559-3},
	doi = {10.1038/s41586-018-0559-3},
	number = {7728},
	urldate = {2026-05-17},
	journal = {Nature},
	publisher = {Nature Publishing Group},
	author = {Liu, Yang and Zhao, Qi and Li, Ming-Han and Guan, Jian-Yu and Zhang, Yanbao and Bai, Bing and Zhang, Weijun and Liu, Wen-Zhao and Wu, Cheng and Yuan, Xiao and Li, Hao and Munro, W. J. and Wang, Zhen and You, Lixing and Zhang, Jun and Ma, Xiongfeng and Fan, Jingyun and Zhang, Qiang and Pan, Jian-Wei},
	month = oct,
	year = {2018},
	pages = {548--551},
}

@article{liu_device-independent_2021,
	title = {Device-independent randomness expansion against quantum side information},
	volume = {17},
	copyright = {2021 The Author(s), under exclusive licence to Springer Nature Limited},
	issn = {1745-2481},
	url = {https://www.nature.com/articles/s41567-020-01147-2},
	doi = {10.1038/s41567-020-01147-2},
	number = {4},
	urldate = {2026-05-17},
	journal = {Nature Physics},
	publisher = {Nature Publishing Group},
	author = {Liu, Wen-Zhao and Li, Ming-Han and Ragy, Sammy and Zhao, Si-Ran and Bai, Bing and Liu, Yang and Brown, Peter J. and Zhang, Jun and Colbeck, Roger and Fan, Jingyun and Zhang, Qiang and Pan, Jian-Wei},
	month = apr,
	year = {2021},
	pages = {448--451},
}

@article{pan_experimental_2022,
	title = {Experimental demonstration of high-rate discrete-modulated continuous-variable quantum key distribution system},
	volume = {47},
	copyright = {© 2022 Optica Publishing Group},
	issn = {1539-4794},
	url = {https://opg.optica.org/ol/abstract.cfm?uri=ol-47-13-3307},
	doi = {10.1364/OL.456978},
	number = {13},
	urldate = {2026-05-23},
	journal = {Optics Letters},
	publisher = {Optica Publishing Group},
	author = {Pan, Yan and Wang, Heng and Shao, Yun and Pi, Yaodi and Li, Yang and Liu, Bin and Huang, Wei and Xu, Bingjie},
	month = jul,
	year = {2022},
	pages = {3307--3310},
}

\newpage
\onecolumngrid
\newpage
\appendix

\section{Semidefinite programs to bound the Shannon entropy}\label{app.sdp_shannon}

In this part of the supplementary material, we present the semidefinite progams (SDPs) we compute the bound the certifiable Shannon entropy in our  protocol. We begin showing the primal SDP directly obtained from Ref.\cite{carceller2025}. Afterwards, we derive dual objective function that will serve as the min-tradeoff function to estimate the non-i.i.d.~randomness certification through the entropy accumulation theorem.

\subsection{Primal SDP}

The Shannon entropy of Bob's measurement outcomes ($H(B|\Lambda_{E})$ in the main text) is lower bounded by \cite{brown2024,carceller2025improving}
\begin{align}
H(B|\Lambda_{E}) \geq c_{m} + \sum_{i=1}^{m-1} \frac{w_i}{t_i\log 2} \inf_{\left\{z_{i,b,x}^{\lambda}\right\}} \left\{ \sum_x p_x \sum_\lambda q(\lambda) \sum_b p_\lambda(b|x)\left[2z_{i,b,x}^\lambda + (1-t_i)(z_{i,b,x}^\lambda)^2\right] + t_i(z_{i,b,x}^\lambda)^2 \right\} \ ,
\end{align}
for $c_m=\sum_i \frac{w_i}{t_i\log 2}$, where $t_i$ and $w_i$ are the nodes and weights of the Gauss-Radau quadrature, $m$ is the quadrature limit, and $\{z_{i,b,x}^\lambda\}$ are arbitrary real variables. Our goal is to minimize the lower-bound on the right-hand side over all possible sate preparations $\{\rho_x^{(\lambda)}\}_x$, POVMs $\{M_b^{(\lambda)}\}_b$ and distributions $q(\lambda)$ such that the semi-DI assumption on the averaged vacuum-component $\omega$ is satisfied, and the observable correlations $W$ is also fulfilled. This is summarised in the optimisation problem,
\begin{align}
c_{m} + \sum_{i=1}^{m-1} \frac{w_i}{t_i\log 2} \quad \underset{\left\{z_{b,x}^{\lambda},q(\lambda),\rho_x^{(\lambda)},M_b^{(\lambda)}\right\}}{\text{minimize}} & \  \left\{ \sum_x p_x \sum_\lambda q(\lambda) \sum_b \tr\left(\rho_x^{(\lambda)}M_b^{(\lambda)}\right)\left[2z_{b,x}^\lambda + (1-t_i)(z_{b,x}^\lambda)^2\right] + t_i(z_{b,x}^\lambda)^2 \right\} \\
\text{such that} \qquad & \quad \rho_x^{(\lambda)}\succeq 0 , \quad \tr(\rho_x^{(\lambda)}) = 1, \\
& \quad M_b^{(\lambda)}\succeq 0 , \quad \sum_b M_b^{(\lambda)}= \mathds{1}, \\
& \quad q(\lambda) \geq 0 , \quad \sum_\lambda q(\lambda) = 1, \\
& \quad \sum_{x}p_x\sum_\lambda q(\lambda)\tr\left(\rho_x^{(\lambda)}\sigma\right)\geq \omega, \\
& \quad \sum_x p_x \sum_\lambda q(\lambda) \tr\left(\rho_x^{(\lambda)}M_x^{(\lambda)}\right) \geq W \ ,
\end{align}
where we defined $\sigma=\ketbra{0}{0}$. Note that we dropped the sub-index $i$ from the arbitrary scalar variable $z_{i,a}^\lambda$, given that the minimisation is performed for each different quadrature element $i$ independently. With the approach introduced in \cite{carceller2025}, we can reformulate this optimisation problem as an SDP relaxation. To do that, we begin listing all relevant operators from our scenario in the lists $L^{\lambda}=\{\rho_x^{(\lambda)},M_b^{(\lambda)},\sigma\}$. Then, we sample monomials from the lists $L^{\lambda}$ up to a desired order $k$ and store them in $S^{\lambda}=\{\mathds{1},\rho_x^{(\lambda)},M_b^{(\lambda)},\sigma,\rho_x^{(\lambda)} M_b^{(\lambda)},\ldots\}$. To this end, we build the moment matrices $\Gamma$, $\Upsilon^{b,x}$ and $\Omega^{b,x}$ defined as
\begin{align}
\Gamma & = \sum_j \ketbra{j_u}{j_v} \Gamma_{u,v}, \quad \text{with} \quad \Gamma_{u,v} = \sum_\lambda q(\lambda) \tr\!\left(u_\lambda^\dagger v_\lambda\right)   \\
\Upsilon^{b,x} & = \sum_j \ketbra{j_u}{j_v} \Upsilon^{b,x}_{u,v}, \quad \text{with} \quad \Upsilon^{b,x}_{u,v} = \sum_\lambda q(\lambda) z_{b,x}^\lambda \tr\!\left(u_\lambda^\dagger v_\lambda\right)   \\
\Omega^{b,x} & = \sum_j \ketbra{j_u}{j_v} \Omega^{b,x}_{u,v}, \quad \text{with} \quad \Omega^{b,x}_{u,v} = \sum_\lambda q(\lambda) (z_{b,x}^\lambda)^2 \tr\!\left(u_\lambda^\dagger v_\lambda\right),
\end{align}
for $u_\lambda,v_\lambda \in S^\lambda$, and where $j_u$ and $j_v$ correspond to the indices of $u_\lambda$ and $v_\lambda$ in $S^\lambda$, respectively. Note that, by definition, the matrices $\Gamma$ and $\Omega^{b,x}$ are positive semidefinite. In addition, one finds that the following block-matrix is also positive semidefinite,
\begin{align}
G^{b,x} = \begin{pmatrix}
\Gamma & \Upsilon^{b,x} \\
\Upsilon^{b,x} & \Omega^{b,x}
\end{pmatrix} =  \sum_\lambda q(\lambda) \sum_j \tr\!\left(u_\lambda^\dagger v_\lambda\right) \ketbra{j_u}{j_v} \otimes  \begin{pmatrix}
1 & z_{b,x}^\lambda \\
z_{b,x}^\lambda & (z_{b,x}^\lambda)^2
\end{pmatrix} \succeq 0 \ .
\end{align}
Our goal is to define an SDP relaxation using only the entries of the moment matrix $G^{b,x}$ as SDP variables. It will be conceptually important to stop and make a remark here. Until now, for the sake of clarity we explicitly preserved the dependence of $\lambda$ of each element of the monomial list. However, observe that all elements in each moment matrix are SDP variables that represent quantities averaged over $\lambda$. Given that the whole problem boils down to optimising over those variables, from now on the dependence of $\lambda$ is removed, as we work directly with the entries of $G^{b,x}$. However, it is important to keep in mind what they represent at the moment of writing down the constraints in our relaxation, as they must faithfully represent the original problem. 

We are now ready to continue properly building the moment matrix $G^{b,x}$, including all the reduction rules derived directly from the quantum theory. In addition to the cyclic invariance of the trace, these include:
\begin{itemize}
\item Purity of quantum states: $\Gamma_{\rho_x,\rho_x} = \Gamma_{\mathds{1},\rho_x}$, $\Upsilon^{b,x'}_{\rho_x,\rho_x} = \Upsilon^{b,x'}_{\mathds{1},\rho_x}$, $\Omega^{b,x'}_{\rho_x,\rho_x} = \Omega^{b,x'}_{\mathds{1},\rho_x}$;
\item Orthogonal projectivity of measurements: $\Gamma_{M_b,M_{b'}} = \delta_{b,b'}\Gamma_{\mathds{1},M_b}$, $\Upsilon^{b'',x}_{M_b,M_{b'}} = \delta_{b,b'}\Upsilon^{b'',x}_{\mathds{1},M_b}$, $\Omega^{b'',x}_{M_b,M_{b'}} = \delta_{b,b'}\Omega^{b'',x}_{\mathds{1},M_b}$;
\item Projectivity of vacuum component: $\Gamma_{\sigma,\sigma} = \Gamma_{\mathds{1},\sigma}$, $\Upsilon^{b,x}_{\sigma,\sigma} = \Upsilon^{b,x}_{\mathds{1},\sigma}$, $\Omega^{b,x}_{\sigma,\sigma} = \Omega^{b,x}_{\mathds{1},\sigma}$;
\end{itemize}
Once the moment matrix $G^{b,x}$ is properly built, we can impose further constraints such as the normalisation of quantum states and the trace-one of the vacuum-component with
\begin{align}
\Gamma_{\mathds{1},\rho_x} = \Gamma_{\mathds{1},\sigma} = 1, \quad \Upsilon^{b,x'}_{\mathds{1},\rho_x} = \Upsilon^{b,x'}_{\mathds{1},\sigma}, \quad \Omega^{b,x'}_{\mathds{1},\rho_x} = \Omega^{b,x'}_{\mathds{1},\sigma}.
\end{align}
Additionally, we impose the specific constraints in our optimisation problem. These are the bounded averaged vacuum component and the correlation limit respectively through
\begin{align}
\sum_x p_x \Gamma_{\rho_x,\sigma} \geq \omega, \quad \text{and} \quad \sum_x p_x \Gamma_{\rho_x,M_x} \geq W .
\end{align}
Finally, from the elements in the moment matrix $G^{b,x}$ we can recover the objective function of the original optimisation problem. With all that, we are able to minimize the Shannon entropy in our problem through the following series of SDP relaxations,
\begin{subequations}
\label{eq:primal}
\begin{align}
c_{m} + \sum_{i=1}^{m-1} \frac{w_i}{t_i\log 2} \quad \underset{\{G^{b,x}\}}{\text{minimize}} & \quad  \left\{ \sum_x p_x \sum_b \left[2\Upsilon^{b,x}_{\rho_x,M_b} + (1-t_i)\Omega^{b,x}_{\rho_x,M_b}\right] + t_i\Omega^{b,x}_{\rho_x,\mathds{1}} \right\} \label{eq:primal_obj} \\
\text{such that} & \quad G^{b,x} = \begin{pmatrix}
\Gamma & \Upsilon^{b,x}  \\
\Upsilon^{b,x} & \Omega^{b,x}
\end{pmatrix} \succeq 0 \ \forall b,x \ , \label{eq:primal_ct1} \\
& \quad \Gamma_{\mathds{1},\rho_x} = 1, \ \forall x \ , \label{eq:primal_ct2} \\
& \quad \Gamma_{\mathds{1},\sigma} = 1, \ , \label{eq:primal_ct3} \\
& \quad \Upsilon^{b,x'}_{\mathds{1},\rho_x} = \Upsilon^{b,x'}_{\mathds{1},\sigma}, \ \forall b,x,x' \ , \label{eq:primal_ct4} \\
& \quad \Omega^{b,x'}_{\mathds{1},\rho_x} = \Omega^{b,x'}_{\mathds{1},\sigma}, \ \forall b,x,x' \ , \label{eq:primal_ct5} \\
& \quad \sum_x p_x \Gamma_{\rho_x,\sigma} \geq \omega, \label{eq:primal_ct6}  \\
& \quad \sum_x p_x \Gamma_{\rho_x,M_x} \geq W . \label{eq:primal_ct7}
\end{align}
\end{subequations}
All the correlations obtained from the moment matrix $G^{b,x}$ are drawn from an outer approximation of the quantum set which grows closer to the true quantum set as we increase the order $k$ of the included monomials. This means that the result obtained from the SDP above will represent a valid lower-bound on the Shannon entropy will grows closer to the true lower-bound (permitted by the Gauss-Radau quadrature) as we increase the level of the relaxation determined by $k$.

\subsection{Dual SDP and min-tradeoff function}

We solve the primal SDP in \eqref{eq:primal} using the solver MOSEK with CVXPY in Python. Since CVXPY internally constructs and solves both the primal and the dual problems, the dual variables associated with each constraint are directly accessible at optimality. Therefore, to reconstruct the dual objective function of the primal SDP, it suffices to identify which primal constraints contribute to the dual objective. To do that, recall that only those constraints containing parameters that are not optimization variables in the primal formulation contribute to the dual objective function. Based on this observation, and for each quadrature iteration $i$, we identify the constraints \eqref{eq:primal_ct2}, \eqref{eq:primal_ct3}, \eqref{eq:primal_ct6}, and \eqref{eq:primal_ct7}. We associate to these constraints the dual variables $r_{i,x}$, $s_{i}$, $v_i$ and $l_i$ respectively, evaluated at their optimal values. By solving the primal SDP with CVXPY and extracting these optimal dual variables, we can explicitly construct the complete dual objective function. This allows us to define the min-tradeoff function as
\begin{align}
f_{\min}(W,\omega) = c_m + \sum_{i=1}^{m-1} \frac{w_i}{t_i\log 2} \left( -\sum_x r_{i,x} - s_i + v_i W + l_i \omega  \right) = \xi + \beta W + \gamma \omega \ ,
\end{align}
where we defined the parameters in their compact form $\xi = c_m - \sum_{i=1}^{m-1}\frac{w_i}{t_i\log 2}\left(\sum_x r_{i,x} + s_i\right)$, $\beta = \sum_{i=1}^{m-1} \frac{w_i}{t_i\log 2}v_i$ and $\gamma = \sum_{i=1}^{m-1} \frac{w_i}{t_i\log 2}l_i$.

\section{Non-i.i.d.~finite-size corrections}
\label{app.eat}

To estimate the correction due to non-i.i.d.~finite-size effects, we employ the generalized Entropy Accumulation Theorem \cite{metger2022}. This states that, after a $n$ non-i.i.d.~experimental rounds, the smooth conditional min-entropy per-round $\frac{1}{n}H_{\min}^\varepsilon(\mathbf{B}|\mathbf{E})$ is lower-bounded by
\begin{equation}\label{eq:geat}
    \frac{1}{n}H^\epsilon_\text{min} (\mathbf{B}|\mathbf{E}) \geq f_\text{min} (W,\omega) - \frac{\alpha -1}{2 - \alpha} \frac{\ln (2)}{2} V^2 - \frac{1}{n} \frac{g(\epsilon) + \alpha \log (1/ p_\Omega)}{\alpha - 1} - (\frac{\alpha -1 }{2 - \alpha})^2 K'(\alpha).
\end{equation}
Here $p_{\Omega}=\frac{1}{4}$ are the prior probabilities, the total number of samples is $n=52428400$, and we select $\alpha=1+\frac{1}{\sqrt{n}}$ to ensure the desired finite-size correction scaling $\mathcal{O}(\frac{1}{\sqrt{n}})$. The other quantities are
\begin{align}
    g(\epsilon) &= - \log(1- \sqrt{1 - \epsilon^2}),\\
    V &= \log (2 n_B^2 + 1) + \sqrt{2 + \text{Var} (f_\text{min})},\\
    K'(\alpha) &= \frac{(2-\alpha)^3}{6(3-2\alpha)^3 \ln (2)} 2^{\frac{\alpha-1}{2-\alpha} (2 \log n_B + \text{Max} (f_\text{min}) - \text{Min} (f_\text{min}))} \ln^3 (2^{2 \log n_B + \text{Max} (f_\text{min}) - \text{Min} (f_\text{min})} + e^2),
\end{align}
where $\epsilon=10^{-4}$ is the smoothing parameter, $n_B=4$ is the number of measurement outcomes and $\text{Max} (f_\text{min})$, $\text{Min} (f_\text{min})$ and $\text{Var} (f_\text{min})$ represent the maximum, the minimum and the variance of the min-tradeoff function $f_\text{min}$, respectively. Since $f_\text{min}(W,\omega)$ is a linear function, these quantities are simply calculated as
\begin{align}
    \text{Max} (f_\text{min}) &= \max_{W,\omega}\left\{f_\text{min}(W,\omega)\right\} \ , \\
    \text{Min} (f_\text{min}) &= \min_{W,\omega}\left\{f_\text{min}(W,\omega)\right\} \ , \\
    \text{Var} (f_\text{min}) &= \gamma^2 \text{Var}(\omega) + \beta^{2} \text{Var}(W) \ .
\end{align}
For each of the $10$ collected data-points we estimate the maximum, minimum and variance of $W$ and $\omega$. Altogether renders a lower-bound on the smooth conditional min-entropy.

\section{Semidefinite program to bound the min-entropy}\label{app.sdp_min}

In this section, we show the semidefinite program (SDP) used in the main text to bound the min-entropy in our semi-device-independent framework assuming a bounded vacuum-component averaged over all state preparations. This method is directly extracted from Ref.~\cite{carceller2025}.

The min-entropy of the measurement outcomes with respect to a classically-correlated eavesdropper in a prepare-and-measure scenario can be written as 
\begin{align}
H_{\min}(B|\Lambda_E)=-\log_2(p_g) , \quad \text{where} , \quad p_g=\sum_x p_x \sum_\lambda q(\lambda) \max_b\{p_\lambda(b|x)\}
\end{align}
is the eavesdropping guessing probability and $p_\lambda(b|x)=\tr\left(\rho_x^{(\lambda)}M_b^{(\lambda)}\right)$ are the correlations per-round achievable with the implementation strategy $\lambda$. For lower-bounding the min-entropy therefore, it is enough to find an upper bound on the guessing probability. In our case, this reduces to the optimisation problem,
\begin{align}
\underset{\{q(\lambda),\rho_x^{(\lambda)},M_b^{(\lambda)}\}}{\text{maximize}} & \quad \sum_x p_x \sum_\lambda q(\lambda) \max_b\left\{\tr\left(\rho_x^{(\lambda)}M_b^{(\lambda)}\right)\right\} \\
\text{such that} \qquad & \quad \rho_x^{(\lambda)}\succeq 0 , \quad \tr(\rho_x^{(\lambda)}) = 1, \\
& \quad M_b^{(\lambda)}\succeq 0 , \quad \sum_b M_b^{(\lambda)}= \mathds{1}, \\
& \quad q(\lambda) \geq 0 , \quad \sum_\lambda q(\lambda) = 1, \\
& \quad \sum_{x}p_x\sum_\lambda q(\lambda)\tr\left(\rho_x^{(\lambda)}\sigma\right)\geq \omega, \\
& \quad \sum_x p_x \sum_\lambda q(\lambda) \tr\left(\rho_x^{(\lambda)}M_x^{(\lambda)}\right) \geq W \ .
\end{align}
where we denoted $\sigma=\ketbra{0}{0}$. In order to cast this optimisation as an SDP relaxation, we firstly need to address the $max_b$ in the objective function. To do that, we invoke Proposition 1 from \cite{Bancal2014}, and reduce the number of considered strategies $\lambda$ to the relevant ones in our case. Namely, we define the tuple $\vec{\lambda} \equiv (\lambda_0,\lambda_1,\lambda_2,\lambda_3)$ to label the strategy which, for input $x$, yields the maximal correlation for outcome $b=\lambda_x$. With that, the guessing probability simply reads $p_g=\sum_x p_x \sum_{\vec{\lambda}} q(\vec{\lambda}) \tr\left(\rho_x^{\vec{\lambda}}M_{\lambda_x}^{\vec{\lambda}}\right)$.

Next, we use the approach introduced in \cite{carceller2025} and proceed defining the SDP relaxation. We define the list of relevant operators $L^{\vec{\lambda}}=\{\rho_x^{\vec{\lambda}},M_b^{\vec{\lambda}},\sigma\}$. Then, we sample monomials from the lists $L^{\vec{\lambda}}$ up to a desired order $k$ and store them in $S^{\vec{\lambda}}=\{\mathds{1},\rho_x^{\vec{\lambda}},M_b^{\vec{\lambda}},\sigma,\rho_x^{\vec{\lambda}} M_b^{\vec{\lambda}},\ldots\}$. To this end, we build the matrix 
\begin{align}
\Gamma^{\vec{\lambda}} & = \sum_j \ketbra{j_u}{j_v} \Gamma^{\vec{\lambda}}_{u,v}, \quad \text{with} \quad \Gamma^{\vec{\lambda}}_{u,v} = q(\vec{\lambda}) \tr\!\left(u_{\vec{\lambda}}^\dagger v_{\vec{\lambda}}\right) \ ,
\end{align}
for $u_{\vec{\lambda}},v_{\vec{\lambda}}\in L^{\vec{\lambda}}$, and where $j_u$ and $j_v$ correspond to the indices of $u_{\vec{\lambda}}$ and $v_{\vec{\lambda}}$ in $S^{\vec{\lambda}}$, respectively. Each element in $\Gamma^{\vec{\lambda}}$ represents a moment build form the relevant operators in our scenario, and by definition it is positive semidefinite.

Some of the properties form the operators defining the moments in $\Gamma^{\vec{\lambda}}$ are inherited into its structure. Concretely, in addition to the cyclicity of the trace, we have: i) the purity of quantum states $\Gamma^{\vec{\lambda}}_{\rho_x,\rho_x}=\Gamma^{\vec{\lambda}}_{\mathds{1},\rho_x}$; ii) orthogonality of projective measurements $\Gamma^{\vec{\lambda}}_{M_b,M_{b'}}=\Gamma^{\vec{\lambda}}_{\mathds{1},M_b}\delta_{b,b'}$; and iii) projectivity of the vacuum $\Gamma^{\vec{\lambda}}_{\sigma,\sigma}=\Gamma^{\vec{\lambda}}_{\sigma,\mathds{1}}$.

Once the moment matrices are propperly built, we can impose additional linear constraints, as for example the normalisation of quantum states and trace-one vacuum-component projector with $\sum_{\vec{\lambda}}\Gamma^{\vec{\lambda}}_{\mathds{1},\rho_x}=\sum_{\vec{\lambda}}\Gamma^{\vec{\lambda}}_{\mathds{1},\sigma}=1$. Finally, we can respectively impose the fulfilment of the SDI assumption on the vacuum component, the observed witness as, and write the guessing probability as
\begin{align}
\sum_x p_x \sum_{\lambda} \Gamma_{\rho_x,\sigma}^{\vec{\lambda}} \geq \omega, \quad \text{and} \quad \sum_x p_x \sum_{\lambda} \Gamma_{\rho_x,M_x}^{\vec{\lambda}} \geq W , \quad p_g=\sum_x p_x \sum_{\vec{\lambda}} \Gamma^{\vec{\lambda}}_{\rho_x^,M_{\lambda_x}} \ .
\end{align}
With all that, we can fully rewrite the original optimisation problem using only the moment matrix as 
\begin{subequations}
\label{eq:primal}
\begin{align}
\underset{\{\Gamma^{\vec{\lambda}}\}}{\text{maximize}} & \quad  \sum_x p_x \sum_{\vec{\lambda}} \Gamma^{\vec{\lambda}}_{\rho_x^,M_{\lambda_x}} \label{eq:primal_obj} \\
\text{such that} & \quad \Gamma^{\vec{\lambda}} \succeq 0, \ \forall \vec{\lambda} \ , \label{eq:primal_ct1} \\
& \quad \Gamma^{\vec{\lambda}}_{\mathds{1},\rho_x} = 1, \ \forall x, \vec{\lambda} \ , \label{eq:primal_ct2} \\
& \quad \Gamma^{\vec{\lambda}}_{\mathds{1},\sigma} = 1, \ \forall \vec{\lambda} , \label{eq:primal_ct3} \\
& \quad \sum_x p_x \Gamma^{\vec{\lambda}}_{\rho_x,\sigma} \geq \omega, \label{eq:primal_ct6}  \\
& \quad \sum_x p_x \Gamma^{\vec{\lambda}}_{\rho_x,M_x} \geq W . \label{eq:primal_ct7}
\end{align}
\end{subequations}
All the correlations obtained from the moment matrices $\Gamma^{\vec{\lambda}}$ are drawn from an outer approximation of the quantum set which grows closer to the true quantum set as we increase the order $k$ of the included monomials. This means that the result obtained from the SDP above will represent a valid upper-bound on the guessing probability, and consequently a valid lower-bound on the min-entropy, which grows closer to the true lower-bound as we increase the level of the relaxation determined by $k$.

\section{Methods}
Each acquired trace is processed with a \SI{50}{\mega\hertz} \glsxtrlong{hpf} to suppress low-frequency technical noise, followed by a \gls{rrc} matched filter with roll-off factor \(0.3\) to minimize inter-symbol interference. 
The phase drift is compensated by applying a global phase rotation to the measured samples, so that the constellation is aligned with the nominal \gls{qpsk} quadrants.
The experimental conditional probabilities $p(b|x)$ are then estimated as relative frequencies by assigning each sample to one of the four phase-space quadrants (outcomes $b$), conditioned on the prepared state $x$.

For each attenuation setting, the mean photon number of the prepared coherent states was inferred from an optical power measurement performed with a calibrated powermeter on a tapped output of the signal line.
Denoting by $P_{\mathrm{opt}}$ the measured optical power at the tapping port, the mean photon number per symbol at the receiver input is estimated as
\begin{equation}
|\alpha|^2 \equiv \langle n\rangle
= \frac{P_{\mathrm{opt}}\,\kappa}{E_{\mathrm{ph}}\,R}\, ,
\label{eq:alpha2_from_power}
\end{equation}
where $R$ is the symbol rate, $E_{\mathrm{ph}}=hc/\lambda$ is the photon energy at wavelength $\lambda$, and $\kappa$ accounts for the known power ratio factor between the tapping port and the receiver input.

\end{document}